\pdfoutput=1
\documentclass[11pt]{article}
\usepackage{jheppub}
\usepackage{graphicx,color} 
\usepackage{amsmath}
\usepackage{amssymb}
\usepackage{comment}
\usepackage{multirow}
\usepackage{mathtools}
\usepackage{dsdshorthand}
\usepackage{shuffle}
\usepackage{makecell}

\newcommand{\bea}{\begin{equation}\begin{aligned}}
\newcommand{\eea}[1]{\label{#1}\end{aligned}\end{equation}}
\newcommand{\beq}{\begin{equation}}
\newcommand{\eeq}{\end{equation}}

\newcommand   \dd  {\mathrm{d}}
\newcommand   \ii  {\mathrm{i}}

\newcommand{\es}[2] {\begin{equation} \label{#1} \begin{split} #2 \end{split} \end{equation}}

\usepackage{tikz}
\usetikzlibrary{arrows,calc,shapes,decorations.pathmorphing,decorations.markings}
\tikzset{
>=stealth',
help lines/.style={dashed, thick},
axis/.style={<->},
important line/.style={thick},
connection/.style={thick, dotted},
  cross/.style={
    cross out,
    draw=black, 
    minimum size=5pt, 
    inner sep=0pt,
    outer sep=0pt
  },
->-/.style={decoration={
  markings,
  mark=at position #1 with {\arrow{>}}},postaction={decorate}}
}

\title{The AdS Veneziano amplitude at small curvature}
\author[a]{Luis F. Alday,}
\author[b]{Shai M. Chester,}
\author[a,c]{Tobias Hansen}
\author[b]{and De-liang Zhong}

\affiliation[a]{Mathematical Institute, University of Oxford,
Woodstock Road, Oxford, OX2 6GG, UK}
\affiliation[b]{Blackett Laboratory, Imperial College, Prince Consort Road, London, SW7 2AZ, UK}
\affiliation[c]{Department of Mathematical Sciences, Durham University,
Stockton Road, Durham, DH1 3LE, UK}
\emailAdd{luis.alday@maths.ox.ac.uk}
\emailAdd{s.chester@imperial.ac.uk}
\emailAdd{tobias.p.hansen@durham.ac.uk}
\emailAdd{d.zhong23@imperial.ac.uk}

\abstract{
We compute the AdS Veneziano amplitude for type IIB gluon scattering in $AdS_5 \times S^3$ to all orders in $\alpha'$ in a small curvature expansion. This is achieved by combining a dispersion relation in the dual $4d$ $\mathcal{N}=2$ SCFT with an ansatz for the amplitude as a worldsheet integral in terms of multiple polylogarithms. The first curvature correction is fully fixed in this way and satisfies consistency checks in the high energy limit, the low energy expansion as previously fixed using supersymmetric localisation, and for the energy of massive string operators, which we independently compute using a semiclassical expansion. We also combine localisation with this first curvature correction to fix the unprotected $D^4F^4$ correction to the amplitude at finite curvature.
}

\begin{document}
\maketitle

\section{Introduction}

The worldsheet is the best understood definition of string theory. It allows us to in principle compute scattering amplitudes in flat spacetime to any order in the string coupling $g_s$ at finite string length $\ell_s=\sqrt{\alpha'}$. For instance, the scattering of four gravitons in type IIB superstring theory at leading order in $g_s$ is known as the Virasoro-Shapiro amplitude, and takes the form of the worldsheet integral (up to an overall factor)
\es{flat_space_VS}{
 A^{(0)}_\text{closed}(S,T)&= \frac{16}{(S+T)^2} \int d^2z \, |z|^{-\frac{S}{2}-2} |1-z|^{-\frac{T}{2}-2}=-\frac{\Gamma(-\frac{S}{4}) \Gamma(-\frac{T}{4})\Gamma(-\frac{U}{4})}{\Gamma(\frac{S}{4}+1)\Gamma(\frac{T}{4}+1)\Gamma(\frac{U}{4}+1)}\,,
}
where we define the Mandelstam variables in terms of the momenta $p_i$ as
\beq
\begin{gathered}
S = - \a' (p_1+p_2)^2\,, \qquad
T = - \a' (p_1+p_4)^2\,, \qquad
U = - \a' (p_1+p_3)^2\,.
\label{mandelstams}
\end{gathered}
\eeq
An outstanding question is how to generalise these worldsheet calculations to curved spacetime, such as appears in the AdS/CFT correspondence. The difficulty is that these curved spacetimes typically have finite Ramond-Ramond (RR) flux, which cannot be handled by the traditional Ramond–Neveu–Schwarz (RNS) formalism \cite{Polchinski:1998rr}. While some progress has been made using the Green-Schwarz \cite{Green:1980zg} and pure spinor \cite{Berkovits:2000fe} formalisms, to date no scattering amplitude in curved spacetime with finite RR flux has been computed directly from the worldsheet.

Recent progress has been made for scattering of gravitons in type IIB string theory on $AdS_5\times S^5$ with radius $R$, using the holographic duality to $\mathcal{N}=4$ $SU(N)$ super-Yang-Mills (SYM) with the 't Hooft coupling $\lambda\equiv g_\text{YM}^2N$  related to the string parameters as $R^4/\ell_s^4=\lambda$. The genus-zero amplitude in $AdS_5\times S^5$ is dual to a four-point function of stress tensors in SYM in the leading large $N$ limit to all orders in $1/\lambda$. This holographic correlator is fixed by the analytic bootstrap \cite{Rastelli:2017udc,Alday:2014tsa,Goncalves:2014ffa} to be 
\es{Mintro}{
M_\text{closed}(s,t)=&\frac{1}{N^2}\Big[ \frac{1}{stu}+\frac{\tl\alpha^{(0)}_{R^4}}{\lambda^{3/2}}+\frac{\tl\alpha^{(1)}_{D^2R^4}}{\lambda^{2}}+\frac{\tl\alpha^{(0)}_{D^4R^4}(s^2+t^2+u^2)+\tl\alpha^{(2)}_{D^4R^4}}{\lambda^{5/2}} \\
&\quad +\frac{\tl\alpha^{(0)}_{D^6R^4}stu+\tl\alpha^{(1)}_{D^6R^4}(s^2+t^2+u^2)+\tl\alpha^{(3)}_{D^6R^4}}{\lambda^{3}} +O(\lambda^{-7/2})\Big] +O(N^{-4})\,,
}
which we wrote in terms of the Mellin space variables $s,t,u=-s-t$ that are related by the Mellin transform to the usual cross ratios $U,V$ in position space \cite{Mack:2009mi,Fitzpatrick:2011ia}. The coefficients $\tl\alpha^{(k)}_{\#}$ correspond to higher derivative corrections to supergravity such as $R^4$, and the protected terms shown here were computed using supersymmetric localisation in \cite{Binder:2019jwn,Chester:2020dja,Chester:2019pvm}\footnote{These higher derivative corrections were also computed at finite complex coupling $\tau$ in \cite{Chester:2019jas,Chester:2020vyz}.}. We can then take the flat space limit \cite{Penedones:2010ue} by rescaling $(s,t,u)\to R^2(s,t,u)$ and taking $R/\ell_s\to\infty$, after which we identify the rescaled $s,t,u$ with the flat space Mandelstam variables in \eqref{mandelstams} divided by $\alpha'$. The $1/\lambda$ expansion at finite $s,t,u$ in \eqref{Mintro} then corresponds to expanding the AdS Virasoro-Shapiro amplitude at small $S,T,U$.

We can also consider the AdS Virasoro-Shapiro amplitude at finite $S,T,U$. This limit corresponds to rescaling  $M_\text{closed}(s,t)\to \lambda^{3/2}M_\text{closed}(\sqrt{\lambda}s,\sqrt{\lambda}t)$, and then expanding in large $R/\ell_s=\lambda^{1/4}$,
resulting in a small curvature expansion for the AdS Virasoro-Shapiro
\beq
A_\text{closed}(S,T) = A_\text{closed}^{(0)}(S,T) + \frac{1}{\sqrt{\lambda}} A_\text{closed}^{(1)}(S,T) + O \left(1/\lambda \right)\,,
\label{Aclosed_full}
\eeq
where the leading term is the flat space Virasoro-Shapiro amplitude \eqref{flat_space_VS} and determines an infinite number of coefficients $\tl\alpha^{(0)}_{\#}$ in \eqref{Mintro}, $A_\text{closed}^{(1)}(S,T)$ is the first curvature correction which determines the coefficients $\tl\alpha^{(1)}_{\#}$, and so on.
In \cite{Alday:2022uxp,Alday:2022xwz,Alday:2023jdk,Alday:2023mvu}, it was shown how to compute these curvature corrections using two constraints. The first constraint uses a dispersion relation to relate the correlator to single trace massive string operators that scale as $\lambda^{1/4}$, such as the Konishi operator. The CFT data of these operators can be computed at leading large $\lambda$ from the flat space amplitude, while $1/\lambda$ corrections are computable from integrability applied to the classical string theory on $AdS_5\times S^5$ \cite{Gromov:2011de}. The second constraint comes from assuming that the AdS amplitude can be computed from a worldsheet integral such as \eqref{flat_space_VS}, with the additional insertion of single-valued multiple polylogarithms, which was motivated from the fact that only such functions appear when performing closed string worldsheet calculations in an expansion around flat space. By combining both constraints, the first two curvature corrections were computed in \cite{Alday:2023mvu}, and higher curvature corrections are expected to be fixed in terms of $1/\lambda$ corrections to massive string operators as computable from integrability.
Each curvature correction to the worldsheet integrand turns out to admit uniform transcendentality, a feature that is typical for $\mathcal{N}=4$ SYM.

In this paper, we will generalise this strategy to the scattering of open string gluons on the worldvolume of D7 branes in type IIB string theory. In particular, consider $N$ D3 branes, 4 D7 branes, and an O7 plane in type IIB string theory \cite{Sen:1996vd,Banks:1996nj,Douglas:1996js}. At large $N$, the geometry is $AdS_5\times S^5/\mathbb{Z}_2$, where the $\mathbb{Z}_2$ orientifold has a fixed point locus of $S^3$, such that gluons scattering on the D7 branes probe $AdS_5\times S^3$ \cite{Aharony:1998xz}. The dual CFT is a 4d $\mathcal{N}=2$ $USp(2N)$ gauge theory with one hypermultiplet in the antisymmetric and four hypermultiplets in the fundamental, where the latter transform under an $SO(8)$ flavour symmetry. Gluon scattering on the D7 branes is dual to the four-point function of flavour multiplets in the CFT, such that we can define the AdS Veneziano amplitude as the leading large $N$ limit of this correlator to all orders in the 't Hooft coupling $\lambda$, which is proportional to $R^4/\ell_s^4$. We can similarly consider two different orbifolds of this theory as described in \cite{Ennes:2000fu}, such that AdS Veneziano amplitudes of all three cases are proportional to each other, except that the flavour group $G_F$ now equals either $U(4)$ or $SO(4)\times SO(4)$ instead of $SO(8)$.\footnote{The $G_F=U(4)$ theory has an $SU(N)$ gauge group with two antisymmetric hypermultiplets and four fundamental hypermultiplets, while the $G_F=SO(4)\times SO(4)$ theory has an $USp(N)\times USp(N)$ gauge group with one bifundamental hypermultiplet and four fundamental hypermultiplets. }

The holographic gluon correlator is fixed by the analytic bootstrap \cite{Alday:2021odx} to be 
\begin{align}
{}&M^{I_1 I_2 I_3 I_4}(s,t)=\frac{\Tr(T^{I_1}T^{I_2}T^{I_3}T^{I_4})}{N}\Big[ -\frac{2}{st}+\frac{\tl\alpha^{(0)}_{F^4}}{\lambda}+\frac{\tl\alpha^{(0)}_{D^2F^4} u + \tl\alpha^{(1)}_{D^2F^4}}{\lambda^{3/2}} \label{MintroOpen}\\
&\quad +\frac{\tl\alpha^{(0)}_{D^4F^4,1} u^2 + \tl\alpha^{(0)}_{D^4F^4,2} s t+
\tl\alpha^{(1)}_{D^4F^4} u + \tl\alpha^{(2)}_{D^4F^4}}{\lambda^{2}} +O(\lambda^{-5/2})\Big]+\text{permutations} +O(N^{-2})\,,\nonumber
\end{align}
where we sum over two permutations of the external operators and $T^I$ are generators in the adjoint of the gauge group.
The coefficients $\tl\alpha^{(k)}_{\#}$ correspond to higher derivative corrections to gauge theory such as $F^4$, and the terms in the first line are protected and were computed using supersymmetric localisation in \cite{Behan:2023fqq}\footnote{These first two corrections were also computed at finite complex coupling $\tau$ in \cite{Behan:2023fqq}.}. 
The coefficients $\tl\alpha^{(0)}_{\#}$ are determined via the flat space limit by the flat space scattering amplitude for four gluons in open superstring theory at leading order in $g_s$, known as the Veneziano amplitude
\es{flat_space_amplitude}{
A^{I_1 I_2 I_3 I_4}(p_i)&=g_s^2\hat K\left[\Tr(T^{I_1}T^{I_2}T^{I_3}T^{I_4}) A^{(0)}(S,T)+\text{permutations} \right]\,,\\
 A^{(0)}(S,T)&=\frac{1}{S+T} \int_0^1 dz \, z^{-S-1} (1-z)^{-T-1}=-\frac{\Gamma(-S) \Gamma(-T)}{\Gamma(1-S-T)}\,,
}
where $\hat K$ is a kinematic factor.
In AdS this Veneziano amplitude receives curvature corrections that are suppressed by powers of $1/\sqrt{\lambda}$
\beq
A(S,T) = A^{(0)}(S,T) + \frac{1}{\sqrt{\lambda}} A^{(1)}(S,T) + O \left(1/\lambda \right)\,.
\eeq
The term $A^{(1)}(S,T)$ is the first curvature correction and determines an infinite number of coefficients $\tl\alpha^{(1)}_{\#}$ in \eqref{MintroOpen}.
As in the $\mathcal{N}=4$ SYM case, we can compute $A^{(1)}(S,T)$ by combining a dispersion relation that relates the correlator to single trace massive string operators that scale as $\lambda^{1/4}$, as well as by assuming the correlator is given by a worldsheet integral. Since open string scattering is not single-valued, unlike closed string scattering, we assume a more general ansatz of multiple polylogarithms. Nonetheless, we find these two constraints completely fix the first curvature correction of the AdS Veneziano amplitude, which takes the form of the worldsheet integral in \eqref{flat_space_amplitude}
\beq
A^{(1)}(S,T) = \frac{1}{S+T} \int_0^1 dz \, z^{-S-1} (1-z)^{-T-1} G^{(1)}(S,T,z)\,,
\label{VenInt1}
\eeq
with the extra insertion
\begin{align}
{}&G^{(1)} (S,T,z) =
(S+T)^{-1}\big[3 +(4 T-S)\log (z) +(4 S-T)\log (1-z) -S (3 S+4 T) \log ^2(z) \nonumber\\
&-\hspace{-1pt}T (3T+4S)\log ^2(1-z)
+ \left(5 S^2+12 S T+5 T^2\right)\left(\zeta (2) - \text{Li}_2(z) - \text{Li}_2(1-z) \right)\big]+\zeta (3) (S+T)^2\nonumber\\
&+\hspace{-2pt} T(2 S+T) \left(\log (1-z)\text{Li}_2(1-z) -\text{Li}_3(1-z) \right)
+S^2 \log ^2(z)\left(\tfrac23 \log(z)-\log (1-z)\right)
\nonumber\\
&+\hspace{-2pt} S (2 T+S)\left(\log (z)\text{Li}_2(z)-\text{Li}_3(z)\right)
+T^2 \log^2(1-z) \left(\tfrac23 \log(1-z) - \log(z)\right) \,.
\label{G1_Li}
\end{align}
It is not surprising that this result does not admit uniform transcendentality, given that the theory we consider is not maximally supersymmetric.

We have three distinct consistency checks on our answer. Firstly, in the high energy limit we find that our answer takes an exponential form, as was observed for the AdS Virasoro-Shapiro amplitude in \cite{Alday:2023pzu} and that the exponent for open strings is half of the one for closed strings, as argued in \cite{Gross:1989ge}. Secondly, while integrability has not yet been completed for the open string theories we consider,\footnote{See however \cite{Chen:2004mu,Chen:2004yf,Stefanski:2003qr,Berenstein:2002zw} for some first few steps.} we can still compute the scaling dimension of massive string operators in a large $\lambda$ expansion using the semiclassical approach of \cite{Roiban:2011fe}, which gave the correct answer in the $AdS_5\times S^5$ case. In our open string case, we are able to compute the spin dependent terms of the first $1/\sqrt{\lambda}$ correction to the massive string operator scaling dimensions, which matches our answer for the amplitude. Thirdly, 
the small $S,T$ expansion of our result matches the first line of \eqref{MintroOpen}, which was previously computed in the $G_F=SO(8)$ theory using localisation \cite{Behan:2023fqq}.
Furthermore, we can combine our new constraints with those of \cite{Behan:2023fqq} to fix all terms displayed in the second line of \eqref{MintroOpen}. This corresponds to the $D^4F^4$ higher derivative correction to the super-Yang-Mills term $F^2$ that describes the $AdS_5\times S^3$ effective action, and is the first unprotected correction.

The rest of this paper is organised as follows.
In Section \ref{setup} we discuss kinematic constraints from superconformal symmetry on the flavour multiplet correlator, as well as derive the dispersion relation. In Section \ref{world}, we describe our ansatz for the worldsheet integral, which we combine with the dispersion relation to fix the first curvature correction to the AdS Veneziano amplitude.
In Section \ref{highLow}, we consider the high and low energy limits
and extract the OPE data of massive string operators from the amplitude.
In Section \ref{class}, we compute the energies of massive string operators in a semiclassical expansion.
We conclude in Section \ref{conc} with a review of our results and a discussion of future directions. Technical details of the calculations are given in the various Appendices.

\section{Setup}
\label{setup}

\subsection{Correlator}

We consider 4d $\mathcal{N} = 2$ superconformal field theories with R-symmetry group $SU(2)_R \times U(1)_R$ and flavour symmetry $SU(2)_L \times G_F$, where as discussed above $ G_F$ can be $SO(8)$, $U(4)$, or $SO(4)\times SO(4)$.
We will consider the moment map operator $\mathcal{O}^I(x,v)$, which is the superconformal primary of the flavour supermultiplet, and is a Lorentz scalar with dimension $\Delta=2$ in the singlet of $SU(2)_L$, the adjoint of $SU(2)_R$ (with polarisation $v^\alpha$) and also in the adjoint of $G_F$ (with index $I$). Conformal correlators of such operators have been constructed in \cite{Alday:2021odx} which we will follow for the setup.

The four-point function under study
\beq
\langle \cO^{I_1} (x_1,v_1) \cO^{I_2} (x_2,v_2) \cO^{I_3} (x_3,v_3) \cO^{I_4} (x_4,v_4) \rangle
=\frac{(v_1 \cdot v_2)^2 (v_3 \cdot v_4)^2}{x_{12}^4 x_{34}^4} \cG^{I_1 I_2 I_3 I_4}(U,V,\alpha)\,,
\eeq
can be expressed in terms of the cross-ratios
\begin{equation}
U=\frac{x_{12}^2x_{34}^2}{x_{13}^2x_{24}^2}=z\bar{z}\,,\quad V=\frac{x_{14}^2x_{23}^2}{x_{13}^2x_{24}^2}=(1-z)(1-\bar{z})\,, \quad
\alpha=\frac{(v_1\cdot v_3)(v_2\cdot v_4)}{(v_1\cdot v_2)(v_3\cdot v_4)}\,,
\end{equation}
where
\begin{equation}
x_{ij}=x_i-x_j\;,\quad (v_i\cdot v_j)=v_i^{\alpha}v_j^{\beta}\epsilon_{\alpha\beta}\;.
\end{equation}
We can define a reduced correlator $\cH(U,V)$ by considering the solution to the $\mathcal{N} = 2$ superconformal Ward identity \cite{Dolan:2001tt}
\beq
\cG^{I_1 I_2 I_3 I_4}(U,V,\alpha) = \cG^{I_1 I_2 I_3 I_4}_0(U,V,\alpha)+(1-z \alpha)(1-\bar{z}\alpha) \cH^{I_1 I_2 I_3 I_4}(U,V)\,,
\eeq
where $\cG_0$ is the protected part of the correlator, and all nontrivial data is in $\cH(U,V)$.

We will study the Mellin transform of the reduced correlator\footnote{We use shifted Mellin variables compared to \cite{Alday:2021odx}: $(s,t,u)_\text{here} = (s-2,t-2,\tilde{u}-2)_\text{there}$.}
\beq
  \cH^{I_1 I_2 I_3 I_4}(U, V)
   = \int_{-i \infty}^{i \infty} \frac{ds\, dt}{(4 \pi i)^2} U^{\frac s2 + 1} V^{\frac t2 - 1}
    \Gamma \bigg[1 - \frac s2 \bigg]^2 \Gamma \bigg[1 - \frac t2 \bigg]^2 \Gamma \bigg[1 - \frac u2 \bigg]^2
    M^{I_1 I_2 I_3 I_4}(s, t) \,,
\label{mellin}
\eeq
where $u=-s-t$, and crossing symmetry acts on $M^{I_1 I_2 I_3 I_4}(s, t)$ by
\beq
M^{I_1 I_2 I_3 I_4}(s, t) = M^{I_3 I_2 I_1 I_4}(t, s) = M^{I_2 I_1 I_3 I_4}(s, u)\,.
\eeq
In this work we will only consider the leading contribution of order $1/N$ in the large $N$ expansion\footnote{We will drop the overall factor $1/N$ from formulas.}, which corresponds to open string scattering where the worldsheet has the topology of a disc with four insertions at the boundary. The colour structures for this configuration are single traces of the generators $T^I$ of $G_F$ and the amplitude takes the form\footnote{Structures of the form ${\rm Tr} \left(T^{I_1}T^{I_2} \right){\rm Tr}  \left(T^{I_3}T^{I_4} \right)$ and permutations only appear at subleading orders in $1/N$.}\footnote{The normalisation of the generators is not important for our tree level calculation, as it can be absorbed into the overall normalisation of the amplitude.} \cite{Alday:2021odx,Behan:2023fqq}
\bea
M^{I_1 I_2 I_3 I_4}(s, t) ={}& 
{\rm Tr} \left(T^{I_1}T^{I_2}T^{I_3}T^{I_4} \right)
  M(s,t)
+{\rm Tr} \left(T^{I_1}T^{I_4}T^{I_2}T^{I_3} \right)
M(t,u) \\
&+ {\rm Tr} \left(T^{I_1}T^{I_3}T^{I_4}T^{I_2} \right)
M(u,s)
\,,
\eea{trace_basis}
where $M(s,t)$ is called the colour-ordered amplitude and can have only poles in the $s$- and $t$-channels, the only ones consistent with the colour ordering $(1234)$. Crossing symmetry for $M(s,t)$ implies that
\beq
M(s,t) =  M(t,s)\,.
\eeq

We can also expand the reduced correlator in long superconformal blocks as
\beq
\cH^{I_1 I_2 I_3 I_4}(U,V) = \sum_{\tau,\ell} \sum\limits_{r\in {\bf adj} \otimes{\bf adj}} P_r^{I_1 I_2 I_3 I_4} C^2_{\tau,\ell,r} U^{-1} G_{\tau+2,\ell} (U,V)\,,
\label{OPE_expansion}
\eeq
where the labels are the twist $\tau=\Delta-\ell$ and spin $\ell$ of the superconformal primary, the $4d$ conformal blocks are given by
\bea
 G_{\tau,\ell}(U,V) &=\frac{z\bar z}{z-\bar z}(k_{\tau+2\ell}(z)k_{\tau-2}(\bar z)-k_{\tau+2\ell}(\bar z)k_{\tau-2}( z))\,,\\
k_h(z)&= z^{\frac h2}{}_2F_1(h/2,h/2,h,z)\,,
\eea{4dblock}
and $P_r^{I_1 I_2 I_3 I_4}$ is a projector to the irreducible representation $r$ of $G_F$.
The irreps in the $s$-channel will always include the singlet $\bf 1$, the antisymmetric adjoint ${\bf adj}$, and the traceless symmetric $\bf sym$ in terms of adjoint indices. We can use \eqref{trace_basis} together with symmetry properties of conformal blocks under the exchange $1 \leftrightarrow 2$ to conclude that there are only operators in the representations $\mathbf{1}$ and ${\bf sym}$ with even spin and operators in ${\bf adj}$ with odd spin.

The leading term in this Mellin amplitude was computed in \cite{Alday:2021odx}
\beq
M(s,t) =  -\frac{2}{s \, t} + O(\lambda^{-\frac12}) \,,
\label{leading_M}
\eeq
fixing our normalisation of $M(s,t)$.
Apart from the Mellin amplitude we will also study its Borel transform
\beq \label{borel}
A(S,T) \equiv \frac{\lambda}{8} \int_{\kappa-i\infty}^{\kappa+ i \infty} \frac{d\alpha}{2 \pi i} \, e^\alpha \alpha^{-4} 
M \left( \frac{\sqrt{\lambda} S}{2\alpha}, \frac{\sqrt{\lambda} T}{2\alpha} \right)  \,,
\eeq
where the 't Hooft coupling $\lambda$ is related to the AdS radius $R$ and the Regge slope $\alpha'$ via the AdS/CFT dictionary\footnote{For the $G_F=SO(8)$ theory, the 't Hooft coupling is the usual $\lambda=g_\text{YM}^2N$, while for the other two theories there is an extra factor of two.}
\beq
\sqrt{\lambda} = \frac{R^2}{\alpha'} + O\left(\frac{1}{N}\right) \,.
\label{dictionary}
\eeq
We call $A(S,T)$ the AdS amplitude. It is known since \cite{Penedones:2010ue,Fitzpatrick:2011hu} that the transform \eqref{borel} relates AdS/CFT Mellin amplitudes to flat space amplitudes in the limit $R \to \infty$. In our case 
\beq
A(S,T)= \sum_{k=0}^\infty \frac{1}{\lambda^\frac{k}{2}} A^{(k)}(S,T) \,,
\label{fsl}
\eeq
has the flat space Veneziano amplitude in \eqref{flat_space_amplitude} as the leading term. This leads us to identify $S$, $T$ and $U=-S-T$ with the dimensionless Mandelstam variables given in \eqref{mandelstams}.

\subsection{Dispersion relation}

In order to derive a dispersion relation for the colour-ordered Mellin amplitude $M(s,t)$, we use two main ingredients. The first is that the OPE expansion \eqref{OPE_expansion} translates to the statement \cite{Mack:2009mi,Penedones:2019tng} that the Mellin amplitude has simple poles at $s = \tau + 2m - 2$, where $m \in \mathbb{N}_0$ label (super)conformal descendants, with residues given by
\begin{align}
M(s, t) \approx \frac{C^2_{\tau, \ell}  \mathcal{Q}_{\ell, m}^{\tau +2, d=4}(t-2)}{s-\tau - 2m+2},
\end{align}
where the Mack polynomial $\mathcal{Q}_{\ell, m}^{\tau +2, d=4}(t-2)$ is defined in \eqref{curlyQ}, and we removed the irrep label from $C^2_{\tau, \ell}$ because the leading $1/N$ form of the Mellin amplitude \eqref{trace_basis} and crossing symmetry implies that the symmetric irreps $\bf1$ and $\bf sym$ have even spins $\ell$ and identical CFT data, while the antisymmetric irrep $\bf adj$ can be associated simply with odd spins.\footnote{Other irreps will not contribute nontrivially to the dispersion relation, as single trace operators do not appear in those other irreps.} The second ingredient is the assumption that the string theory amplitude has a softer UV behaviour than the corresponding field theory amplitude \eqref{leading_M}, i.e.\ 
we will assume the following bound in the Regge limit
\beq
M(s, t) = o(u^{-1}) \text{ for } u \to \infty \text{ with } t \text{ fixed}, \text{ Re}(t) < 0\,.
\label{boc}
\eeq
With these assumptions we can derive the fixed-$t$ dispersion relation
\beq
M(s,t) = \oint_{u} \frac{du'}{2 \pi i}  \frac{M(-u'-t,t)}{u'-u}
=\sum_{\tau,\ell,m}\frac{ C_{\tau, \ell}^{2} \mathcal{Q}_{\ell, m}^{\tau +2, d=4}(t-2)}{s-\tau - 2m+2}\,.
\label{dr}
\eeq
We expect the exchanged operators to be related to massive open string states in flat space, so that their dimensions should satisfy
\beq
\Delta = m R \left(1+ O\left( \lambda^{-\frac12} \right) \right) = \sqrt{\delta}\lambda^{\frac14} + O\left( \lambda^{-\frac14} \right)\,,
\eeq
where $\delta = 1,2,\ldots$ is the string mass level and the masses $m^2 = \delta/\alpha'$ can be determined from the locations of the poles of the Veneziano amplitude \eqref{flat_space_amplitude}.
 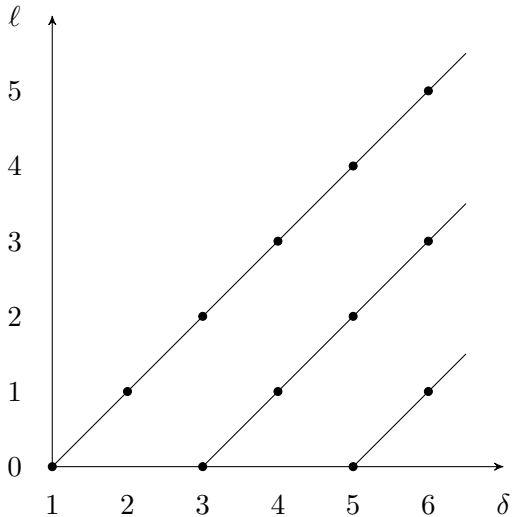
\begin{figure}
\centering
  \begin{tikzpicture}[scale=0.5]
    \coordinate (nw) at (0,12);
    \coordinate (sw) at (0,0);
    \coordinate (se) at (12,0);
    \draw[->] (sw) --  (nw) ;
    \draw[->] (sw) --  (se) ;
    \node at (0,-1) [] {$1$}; 
    \node at (2,-1) [] {$2$}; 
    \node at (4,-1) [] {$3$}; 
    \node at (6,-1) [] {$4$}; 
    \node at (8,-1) [] {$5$}; 
    \node at (10,-1) [] {$6$}; 
    \node at (12,-1) [] {$\delta$}; 
    \node at (-1,0) [] {$0$};
    \node at (-1,2) [] {$1$};
    \node at (-1,4) [] {$2$};
    \node at (-1,6) [] {$3$};
    \node at (-1,8) [] {$4$};
    \node at (-1,10) [] {$5$};
    \node at (-1,12) [] {$\ell$};
    \draw[-] (0,0) --  (11,11) ;
    \draw[-] (4,0) --  (11,7) ;
    \draw[-] (8,0) --  (11,3) ;
    \filldraw [black] (0,0) circle (3pt);
    \filldraw [black] (2,2) circle (3pt);
    \filldraw [black] (4,0) circle (3pt);
    \filldraw [black] (4,4) circle (3pt);
    \filldraw [black] (6,2) circle (3pt);
    \filldraw [black] (6,6) circle (3pt);
    \filldraw [black] (8,0) circle (3pt);
    \filldraw [black] (8,4) circle (3pt);
    \filldraw [black] (8,8) circle (3pt);
    \filldraw [black] (10,2) circle (3pt);
    \filldraw [black] (10,6) circle (3pt);
    \filldraw [black] (10,10) circle (3pt);
  \end{tikzpicture}
\caption{Chew-Frautschi plot of the spectrum of operators exchanged in the Veneziano amplitude.} \label{fig:chew_frautschi}
\end{figure}
The spectrum exchanged in the Veneziano amplitude is shown in figure \ref{fig:chew_frautschi} and the AdS Veneziano amplitude will encode curvature corrections to the flat space masses and partial wave coefficients.
This means we can expand the OPE data as\footnote{We thank Daniele Pavarini for pointing out a typo in a previous version of this equation.}
\bea
\tau(\delta,\ell) &= \sqrt{\delta} \lambda^{\frac{1}{4}} + \tau_1(\delta,\ell)  +  \tau_2(\delta,\ell) \lambda^{-\frac{1}{4}} + \ldots \,,\\
C^2_{\tau,\ell} &=  \frac{2^{-2\tau(\delta,\ell)-2\ell-6} \pi^3 \tau(\delta,\ell)^4}{(\ell+1) \sin(\frac{\pi}{2} \tau(\delta,\ell))^2}  f(\delta,\ell)\,,  \\
 f(\delta,\ell) &= f_0(\delta,\ell) +  f_1(\delta,\ell) \lambda^{-\frac{1}{4}} + f_2(\delta,\ell) \lambda^{-\frac{1}{2}} + \ldots \,,
\eea{ope_data_expansion}
where the flat space spectrum in figure \ref{fig:chew_frautschi} implies
\beq
 f(\delta,\ell) = 0\,, \quad \delta - \ell \text{ even}\,.
\label{flat_spectrum}
\eeq
We can now follow \cite{Alday:2022xwz,Fardelli:2023fyq} to compute the Borel transform \eqref{borel} of the dispersion relation \eqref{dr} in a large $\lambda$ expansion, and to sum over $m$. We find that the amplitude has poles at $S=\delta=1,2,\ldots$. We can match the residues with those of $A^{(0)}(S,T)$ to determine $\langle f_0 \rangle_{\delta, \ell}$ for all exchanged operators,
where the angle brackets indicate a sum over degenerate operators with the same $\delta$ and $\ell$. For instance, we find for the first few Regge trajectories
\bea
\langle f_0 \rangle_{\delta, \delta-1} ={}& r_0(\delta) \,,\\
\langle f_0 \rangle_{\delta, \delta-3} ={}& \frac{r_1(\delta)}{3} \delta(\delta+2)\,,\\
\langle f_0 \rangle_{\delta, \delta-5} ={}& \frac{r_2(\delta)}{90} \delta(5\de^3 + 28 \de^2 + 80 \de + 48)\,,
\eea{f0}
where
\beq
r_n(\delta) = \frac{4^{1-\delta} \delta^{\delta-2n-1} (\delta-2n)^2}{\Gamma(\delta-n+1)}\,.
\label{r}
\eeq
Absence of such poles at order $1/\lambda^{1/4}$ requires that
\begin{align}
\tau_1(\delta, \ell)  = - \ell\,, \qquad \langle f_1 \rangle_{\delta, \ell} = \langle f_0 \rangle_{\delta, \ell} \frac{4 \ell-\frac12}{\sqrt{\delta }}\,.
\label{tau1f1}
\end{align}
The first non-trivial correction to flat space occurs at order $1/\sqrt{\lambda}$, where we compute in an expansion around $S=\delta$
\beq
A^{(1)}(S,T) = \sum\limits_{i=1}^4 \frac{R^{(1)}_i(T,\delta)}{(S-\delta)^i} + O((S-\delta)^0)\,.
\label{A_dr_poles}
\eeq
The numerators $R^{(1)}_i(T,\delta)$ can be explicitly computed in terms of $\langle f_2 \rangle_{\delta, \ell}$ and $\langle f_0 \tau_2 \rangle_{\delta, \ell}$ and are given in appendix \ref{app:summing}, where it is also shown how to obtain them more indirectly by resumming the low energy expansion.

\section{Worldsheet correlator}
\label{world}

In order to fully fix $A^{(1)}(S,T)$ we will now make the assumption that it has a representation as a worldsheet integral. Colour-ordered open string amplitudes can be expressed as an integral over a segment of the boundary of the open string worldsheet. For instance, by conformally mapping the worldsheet to the upper half plane and fixing three of the operator insertions at 0, 1 and $\infty$, the flat space Veneziano amplitude can be written as \eqref{flat_space_amplitude}. Correspondingly, our ansatz for the first curvature correction is
\beq
A^{(1)}(S,T) = \frac{1}{S+T} \int\limits_0^1 dz \, z^{-S-1} (1-z)^{-T-1}
G^{(1)} (S,T,z)\,.
\label{worldsheet_integral}
\eeq
Since $A^{(1)}(S,T)$ has poles up to fourth order \eqref{A_dr_poles}, the integrand $G^{(1)} (S,T,z)$ should have terms with maximal transcendentality 3. Further it should only have singularities at $z=0$ and $z=1$.
In order for the Wilson coefficients in the low energy expansion \eqref{MintroOpen} to have the expected transcendental weight, the weight of each term in $G^{(1)} (S,T,z)$ should match the degree in $S$ and $T$ plus one. Finally, crossing symmetry dictates
\beq
G^{(1)} (S,T,z) = G^{(1)} (T,S,1-z)\,.
\eeq
A suitable basis of transcendental functions are the multiple polylogarithms (MPLs) $L_w(z)$, which are labelled by a word $w$ made up of letters in the alphabet $\{0,1\}$. They can be defined recursively by 
\begin{equation}
\frac{d}{dz} L_{0w}(z) = \frac{1}{z} L_{w}(z)\,,\qquad
\frac{d}{dz} L_{1w}(z) = \frac{1}{z-1} L_{w}(z)\,,
\end{equation}
together with the condition $\lim_{z \to 0} L_w(z)=0$ unless $w=0^p$, for which $L_{0^p}(z) = \frac{\log^p z}{p!}$.
In particular for the empty word we have $L_\emptyset(z)=1$.
This leads us to the ansatz
\beq
G^{(1)} (S,T,z) = \frac{1}{S+T} \sum\limits_{n=0}^{3} \sum_{i,j,\pm} c_{n,i,j}^\pm P^{\pm}_{n,i}(S,T) T^\pm_{n,j}(z)\,,
\eeq
where $P^{\pm}_{n,i}(S,T)$ are symmetric / antisymmetric homogeneous polynomials of degree $n$
and the functions $T^\pm_{n,j}(z)$ have transcendental weight $n$ and are given in terms of
\beq
L^\pm_w (z) = L_w (z) \pm L_w (1-z)\,,
\eeq
by
\bea
T^+_{3}(z) ={}& \left(L^+_{000}(z), L^+_{001}(z), L^+_{010}(z), L^+_{011}(z), \zeta(2) L^+_{0}(z), \zeta(3) \right)\,,\\
T^-_{3}(z) ={}& \left(L^-_{000}(z), L^-_{001}(z), L^-_{010}(z), L^-_{011}(z), \zeta(2) L^-_{0}(z) \right)\,,\\
T^+_{2}(z) ={}& \left(L^+_{00}(z), L^+_{01}(z),  \zeta(2)  \right)\,, \qquad
T^-_{2}(z) = \left(L^-_{00}(z), L^-_{01}(z) \right)\,,\\
T^+_{1}(z) ={}& \left(L^+_{0}(z)  \right)\,, \qquad
T^-_{1}(z) = \left(L^-_{0}(z) \right)\,, \qquad
T^+_{0}(z) = \left(1  \right)\,.
\eea{MPL_list}
The ansatz has 33 rational coefficients $c_{n,i,j}^\pm$.
We start fixing these coefficients by demanding consistency with \eqref{A_dr_poles}, assuming that the OPE data takes the form
\bea
\sqrt{\delta} \langle f_0 \tau_2 \rangle_{\delta, \ell} &= r^{ \tau_2 ,1}_{\delta,\ell} + r^{ \tau_2 ,\zeta(2)}_{\delta,\ell} \zeta(2) \,,\\
\langle f_2 \rangle_{\delta, \ell} &= r^{f_2,1}_{\delta,\ell} + r^{ f_2 ,\zeta(2)}_{\delta,\ell} \zeta(2) + r^{ f_2 ,\zeta(3)}_{\delta,\ell} \zeta(3)\,,
\eea{OPE_assumption}
where $r^{i,j}_{\delta,\ell}$ are rational numbers, together with \eqref{flat_spectrum}.
To compare to \eqref{A_dr_poles} one first Taylor expands the integrand of \eqref{worldsheet_integral} around $z=0$ and then integrates before finally expanding around $S=\delta$, for different values of $\delta$.
This fixes all 33 parameters of the ansatz.
Our solution is
\bea
G^{(1)} (S,T,z) ={}&
\left(S^2+T^2\right) \left(2L^+_{000}(z) -L^+_{011}(z)+\tfrac32 \zeta(3)\right)
+\tfrac{3}{2} L^+_0(z) \\
&-\tfrac{1}{2}  \left(3 S^2+4 S T+3 T^2\right) \left(L^+_{001}(z) + L^+_{010}(z) - \zeta(3) \right)
+\frac{3}{S+T}\\
&-\frac{3 S^2+8 S T+3 T^2}{S+T} L^+_{00}(z)
+\frac{5 S^2+12 S T+5 T^2}{S+T} \left(L^+_{01}(z)+ \zeta (2) \right)\\
&+ (S^2-T^2)\Big(2 L^-_{000}(z)-\tfrac{3}{2} L^-_{001}(z)-\tfrac{3}{2}  L^-_{010}(z)+ L^-_{011}(z)\Big)\\
& - \frac{S-T}{S+T} \Big(3 (S+T) L^-_{00}(z)+\tfrac{5}{2} L^-_0(z)\Big)\,.
\eea{G1}
The same result expressed in terms of classical polylogarithms can be found in \eqref{G1_Li}.

\section{High and low energy limits }
\label{highLow}

Since our solution for the first curvature correction to the AdS Veneziano amplitude relied on various assumptions, it is important to check that it matches other independent calculations. We will first compare to the high energy limit, then the low energy limit as previously computed using localisation for the $G_F=SO(8)$ theory in \cite{Behan:2023fqq}, and finally in the next section to the energies of massive string operators that we compute semiclassically. Note that the first and third checks apply to theories with any $G_F$. In the low energy expansion, we also combine localisation constraints with our new results to completely fix the $1/\lambda^2$ term, which corresponds to $D^4F^4$ correction at finite curvature.

\subsection{High energy limit}

In the high-energy limit of large $S,T,R$ with $S/T$ and $S/R$ fixed\footnote{We set $\alpha'=1$ in this section.} we expect
the amplitude $A(S,T)$ to be determined by a classical computation, as shown for the closed string amplitude on $AdS_5 \times S^5$ in \cite{Alday:2023pzu}.
We expect the form
\beq
A^{\text{HE}}_\text{open}(S,T) \equiv
\lim\limits_{\substack{S,T,R \to \infty\\ S/T, S/R \text{ fixed}}}
A(S,T) \sim
e^{-\mathcal{E}_\text{open}}\,,
\label{HE_limit}
\eeq
and the exponent is determined by the saddle point at $z=\frac{S}{S+T}$ of the integrals \eqref{flat_space_amplitude} and \eqref{worldsheet_integral}
\begin{align}
\mathcal{E}_\text{open} ={}& \mathcal{E}^{(0)}_\text{open} + \frac{1}{R^2} \mathcal{E}^{(1)}_\text{open} + O \left(\frac{1}{S} \right)\,,\nonumber\\
\mathcal{E}^{(0)}_\text{open} ={}& S \log \left(\tfrac{S}{S+T}\right) + T \log \left(\tfrac{T}{S+T}\right)\,,\nonumber\\
\mathcal{E}^{(1)}_\text{open} ={}& - G^{(1)} \left(S,T,\tfrac{S}{S+T}\right)+ O(S)
 \label{Eopen}\\
 ={}&
 S^2 \left(2L_{100} \hspace{-2pt}\left(\tfrac{S}{S+T}\right)
-4 L_{000} \hspace{-2pt}\left(\tfrac{S}{S+T}\right) - \zeta(3)\right)
+S (3 S+2 T) \left( L_{001} \hspace{-2pt}\left(\tfrac{S}{S+T}\right) + L_{010} \hspace{-2pt}\left(\tfrac{S}{S+T}\right) \right)\nonumber\\
&+T^2 \left(2 L_{011} \hspace{-2pt}\left(\tfrac{S}{S+T}\right) - 4L_{111} \hspace{-2pt}\left(\tfrac{S}{S+T}\right) \right)
+T (2 S+3 T) \left( L_{101} \hspace{-2pt}\left(\tfrac{S}{S+T}\right) + L_{110} \hspace{-2pt}\left(\tfrac{S}{S+T}\right) \right)\,.
\nonumber
\end{align}
The exponent can be interpreted as the energy of a classical string solution and it was argued in \cite{Gross:1989ge} that the open string should have half the energy of the corresponding closed string solution, given that gluing two open string worldsheets gives one closed string worldsheet.
To check this, recall that the closed string amplitude on $AdS_5 \times S^5$ takes the form of an integral over the Riemann sphere\footnote{Note that we evaluate \eqref{Aclosed_full} at $4S$, $4T$ because the open string momenta are doubled when applying the reflection principle of \cite{Gross:1989ge}.}
\beq
A_\text{closed}(4S,4T) = \frac{1}{(S+T)^2} \int d^2 z \, |z|^{-2S-2} |1-z|^{-2T-2} \left(
1 + \frac{1}{R^2}G^{(1)}_\text{closed} (S,T,z) + O\left(1/R^4\right)
\right) \,.
\eeq
The high-energy limit is defined as
\beq
A^{\text{HE}}_\text{closed}(4S,4T) \equiv
\lim\limits_{\substack{S,T,R \to \infty\\ S/T, S/R \text{ fixed}}}
A_\text{closed}(4S,4T) \sim
e^{-\mathcal{E}^{(0)}_\text{closed} - \frac{1}{R^2} \mathcal{E}^{(1)}_\text{closed} - O \left(\frac{1}{S} \right)}\,,
\eeq
and one can check that the closed string exponents from \cite{Alday:2023jdk,Alday:2023mvu,Alday:2023pzu} are precisely twice as large as the ones for the open string \eqref{Eopen}
\beq
\mathcal{E}^{(0)}_\text{closed} = 2 \mathcal{E}^{(0)}_\text{open}\,, \qquad
\mathcal{E}^{(1)}_\text{closed} = -G_\text{closed}^{(1)} \left(S,T,\tfrac{S}{S+T}\right) = 2 \mathcal{E}^{(1)}_\text{open}\,.
\eeq
Note that we are comparing different string theories and would not expect such a relation beyond the high-energy limit, which is governed by classical solutions.

\subsection{Low energy expansion}

Let us now discuss the low energy expansion of $A(S,T)$, i.e.\ the Taylor expansion around $S = T = 0$.
We define the Wilson coefficients $\alpha^{(k)}_{a,b}$ by
\beq
A(S,T) =  -\frac{1}{ST}+ \sum\limits_{k,a,b=0}^\infty 
\frac{\hat\sigma_1^a
\hat\sigma_2^b}{\lambda^{\frac{k}{2}}}
 \alpha^{(k)}_{a,b} \,,
\qquad
\hat\sigma_1= -U\,, \quad
\hat\sigma_2 = -ST\,.
\label{LEE}
\eeq
The flat space Wilson coefficients $\alpha^{(0)}_{a,b}$ can be easily extracted from the representation
\beq
A^{(0)}(S,T) = -\frac{1}{S T} \exp \left( \sum_{n=2}^\infty \frac{\zeta(n)}{n} \left( S^n + T^n - (S+T)^n \right)\right) \,,
\eeq
giving for instance
\bea
\alpha^{(0)}_{a,0} ={}& \zeta(a+2)\,,\\
\alpha^{(0)}_{a,1} ={}& \frac12 \sum\limits_{\substack{i_1,i_2=0\\i_1+i_2=a}}^a\zeta(2+i_1) \zeta(2+i_2)  +\frac12 (a+1) \zeta(a+4)\,, \ \ldots
\eea{alpha0_from_flat}
We can compute the low energy expansion of the Mellin amplitude by applying the inverse of \eqref{borel} term by term to \eqref{LEE}, giving
\beq
M(s,t) = -\frac{2}{s \, t} +  \sum\limits_{k,a,b=0}^\infty \frac{\Gamma(4+a+2b)2^{3+a+2b}}{\lambda^{1+\frac{a}{2}+b+\frac{k}{2}}}
\sigma_1^a
\sigma_2^b
 \alpha^{(k)}_{a,b} \,,
\label{wilson_expansion}
\eeq
with
\beq
\sigma_1 = -u\,, \qquad
\sigma_2 = -s t\,.
\eeq
The first few terms of the Mellin amplitude read
\beq
M(s,t) = -\frac{2}{s \, t} + \frac{48 \zeta(2)}{\lambda} - \frac{384 \zeta(3) u-48 \alpha^{(1)}_{0,0}}{\lambda^{\frac32}}  -2^7 3 \frac{\zeta(2)^2 \left(7 s t-4 u^2\right)+ u \alpha^{(1)}_{1,0} -\frac18 \alpha^{(2)}_{0,0}}{\lambda^{2}} + O\left( \lambda^{-\frac52} \right)\,.
\label{melAns}
\eeq
For the $G_F=SO(8)$ theory, the localisation constraint of \cite{Behan:2023fqq} then fixes\footnote{Note that $\alpha^{(1)}_{1,0}$ cancels out in this computation.}
\beq
\alpha^{(1)}_{0,0} = 0\,, \qquad
\alpha^{(2)}_{0,0} = 48 \zeta(2)^2\,,
\label{localisation}
\eeq
where the first coefficient was fixed already in \cite{Behan:2023fqq}, while the second we fix in Appendix \ref{loc}.

To compute the coefficients $\alpha^{(1)}_{a,b}$ from our expression for $A^{(1)}(S,T)$ obtained above, we need to expand integrals of the form
\beq
I_w (S,T) = \int\limits_0^1 dz \, z^{-S-1} (1-z)^{-T-1} L_w (z)\,.
\eeq
This computation is done in appendix \ref{app:LEE} and the result is
\beq
I_w (S,T) = \text{poles } + 
\sum\limits_{p,q=0}^{\infty} (-S)^p (-T)^q 
\sum\limits_{W \in 0^p \shuffle 1^q \shuffle w} 
\left( L_{0W}(1) - L_{1W}(1) \right)\,,
\eeq
where the pole terms can be obtained as explained in the appendix and $\shuffle$ is the shuffle product. Note that $L_{0W}(1)$ and $L_{1W}(1)$ are MZVs of weight $p+q+|w|+1$.
Using this, we can compute the low energy expansion corresponding to \eqref{G1}
\bea
A^{(1)}(S,T) ={}& -3 \hat{\sigma} _1 \zeta (2)^2
+\hat{\sigma} _1^2 \left(\tfrac{19 }{2}\zeta (5)-20 \zeta (2) \zeta (3)\right)
-\hat{\sigma} _2 (19 \zeta (5)+20 \zeta (2) \zeta (3))\\
&-  \hat{\sigma} _1^3 \left(\tfrac{33}{2} \zeta (3)^2 + \tfrac{68}{7} \zeta (2)^3\right)
-\hat{\sigma} _1 \hat{\sigma} _2 \left(\tfrac{33}{2} \zeta(3)^2 + \tfrac{278}{7}  \zeta (2)^3\right)\\
&+\hat{\sigma} _1^4 \left(25 \zeta (7)-51 \zeta (2) \zeta (5)-\tfrac{96}{5} \zeta (2)^2 \zeta (3)\right)\\
&- \hat{\sigma} _1^2 \hat{\sigma} _2 \left(\tfrac{2137 }{16}\zeta (7)+102 \zeta (2) \zeta (5)+\tfrac{288}{5} \zeta (2)^2 \zeta (3)\right)\\
&- \hat{\sigma} _2^2 \left(50 \zeta (7)+51 \zeta (2) \zeta (5)+\tfrac{168}{5} \zeta (2)^2 \zeta (3)\right)+\dots,
\eea{A1_LEE}
where we expanded up to fourth order in $S,T,U$. Note that there are no poles at $S,T=0$ as expected from the field theory Mellin amplitude and we have $\alpha^{(1)}_{0,0}=0$ in agreement with the localisation result \eqref{localisation}.
This also fixes the remaining coefficient at order $1/\lambda^2$ in \eqref{melAns} to
\es{locAns}{
\alpha^{(1)}_{1,0} =-3\zeta(2)^2 \,.
}
We have thus completely determined the $D^4F^4$ correction in AdS at tree level, as given in \eqref{melAns}, \eqref{localisation}, and \eqref{locAns}.
In the notation of \eqref{MintroOpen} we have
\beq
\tl\alpha^{(0)}_{D^4F^4,1} = 2^9 3 \zeta(2)^2\,, \ 
\tl\alpha^{(0)}_{D^4F^4,2} = - 2^7 21 \zeta(2)^2\,, \ 
\tl\alpha^{(1)}_{D^4F^4} = 2^7 9 \zeta(2)^2\,, \ 
\tl\alpha^{(2)}_{D^4F^4} = 2^8 9 \zeta(2)^2\,.
\eeq

\subsection{OPE data}

Let us now extract the OPE data from our solution. From $A^{(1)}(S,T)$ we obtain for the first two Regge trajectories ($r_n(\delta)$ is defined in \eqref{r}, and similar formulas can be obtained for all Regge trajectories)
\bea
\langle f_0 \tau_2 \rangle_{\delta, \delta-1} ={}& r_0(\delta) \sqrt{\delta}\left(\frac{3}{4} \delta +\frac{1}{2\delta }-\frac{3}{4}\right)\,, \\
\langle f_0 \tau_2 \rangle_{\delta, \delta-3} ={}& \frac{r_1(\delta)}{36 } \sqrt{\delta}\left(9 \delta ^3+13 \delta ^2-28 \delta +24\right)\,, \\
\langle f_2 \rangle_{\delta, \delta-1} ={}& \frac{r_0(\delta)}{24 \delta}\left(-14 \delta ^3+198 \delta ^2-580 \delta +243\right) + \delta ^2 \zeta (3) \langle f_0 \rangle_{\delta, \delta-1}  \,, \\
\langle f_2 \rangle_{\delta, \delta-3} ={}& \frac{r_1(\delta)}{216}\left(-42 \delta ^4+ 266 \delta ^3-2696 \delta ^2-2523 \delta +13122\right)
+ \delta ^2 \zeta (3) \langle f_0 \rangle_{\delta, \delta-3}\,.
\eea{A1_OPE_data}
This implies that the non-degenerate operators on the leading Regge trajectory have the dimensions
\beq
\tau_2(\delta,\delta-1) = \sqrt{\delta} \left( \frac{3\delta}{4} + \frac{1}{2\delta} - \frac34 \right)\,.
\label{tau2}
\eeq
In the next section we will compare \eqref{tau2} to the energies of massive string operators as computed from a semiclassical expansion.
To this end we have to look for classical solutions whose charges correspond to operators from the long superconformal multiplet labelled by $\delta$ and $\ell=\delta-1$,
i.e.\ the long multiplet whose primary has Lorentz spin $\delta-1$, dimension $\sqrt{\sqrt{\lambda}\delta}(1+O(\lambda^{-\frac12}))$,
is in the singlet of $SU(2)_L \times SU(2)_R \times U(1)_R$ and in the ${\bf1}$, ${\bf sym}$ or $\bf{adj}$ of $G_F$.

\section{Massive string operators}
\label{class}

We will now compute the dimension of heavy single trace operators that scale as $\lambda^{1/4} $ and compare to the prediction \eqref{tau2} from our solution. In particular, we will consider single trace operators in irreps with two adjoint indices, as they correspond to open strings with Chan-Paton factors on either end. These can have the irreps ${\bf1}$, ${\bf sym}$ and $\bf{adj}$ as discussed above. Our calculation will apply to all three theories $G_F=SO(8)$, $U(4)$, $SO(4)\times SO(4)$ that we consider, as the latter two differ from the former by orbifolds that do no affect the operator we consider.\footnote{The orbifolds affect only the $SU(2)_L$ part of the geometry, under which our operators are invariant \cite{Ennes:2000fu}.} For simplicity, we will thus discuss only $SO(8)$ in what follows.

From the standard AdS/CFT dictionary, finding the dimension of an operator is equivalent to finding the energy of a string solution in $AdS_5 \times S^5$, which scales as $\lambda^{1/4} \sim \frac{R}{\ell_s}$, in terms of the quantum numbers of the state. By considering a semi-classical expansion in the large 't Hooft coupling regime $\lambda \gg 1$, the authors of \cite{Roiban:2009aa, Roiban:2011fe} have successfully found the dimension of the Konishi operator in $\mathcal{N}=4$ SYM perturbatively in $1/\sqrt{\lambda}$. This was later confirmed and extended using integrability of string theory on $AdS_5\times S^5$ \cite{Gromov:2011de}. Since integrability has not yet been worked out for our open string theory, we will instead use the semiclassical expansion.

To be more precise, recall that the operators in $\mathcal{N}=4$ SYM transform under the $PSU(2,2|4)$ group and thus are labelled by the following quantum numbers of the bosonic subgroups,
\beq
({\bf S}_1, {\bf S}_2 | J_1, J_2, J_3) \quad \text{where} \quad {\bf S}_i, J_i \in \mathbb{Z}/2\,.
\eeq
The ${\bf S}_i$ and $J_i$ labels the Lorentz spin and the $SO(6)$ R-charge. They are related to the $SU(2) \times SU(2) \times SU(4)_R$ Dynkin labels, denoted as $[\mathtt{j}, \bar{\mathtt{j}} |q_1, p , q_2]$, in the following way
\beq \label{eqn-DynkinDict}
[\mathtt{j}, \bar{\mathtt{j}} |q_1, p , q_2] \equiv [{\bf S}_1 + {\bf S}_2, {\bf S}_1- {\bf S}_2| J_2-J_3, J_1 - J_2, J_2 + J_3]\, .
\eeq
Here we use the convention of \cite{Cordova:2016emh} where the dimension of the $\mathtt{j}$ irrep is $\mathtt{j}+1$, namely $\mathtt{j} \in \mathbb{Z}_{>0}$.

On the string side, those quantum numbers are naturally identified with the Noether charges of the string solution, see appendix \ref{apd:classical}, \eqref{apd-ConservedCharges} for details. Therefore, once a string solution is found, we can read off the quantum numbers of the corresponding operator. Since we do not know how to quantise string theory on $AdS_5\times S^5$, we cannot solve the spectrum exactly. Instead, one can use a semiclassical approximation to compute energies of states in the strong coupling regime where $\lambda \gg 1$. More precisely, the semiclassical regime is characterised by the condition that all the charges are large
\begin{equation}
   \mathcal{J}_i \equiv \frac{J_i}{\sqrt{\lambda}} = \text{fixed}, \quad \mathcal{S}_i \equiv \frac{{\bf S}_i}{\sqrt{\lambda}} = \text{fixed}, \quad \mathcal{E} \equiv \frac{E}{\sqrt{\lambda}} = \text{fixed}, \quad \lambda \rightarrow \infty \, .
\end{equation}
Assuming for simplicity that there are only two non-vanishing charges, the energy takes the schematic form
\begin{multline}\label{eqn-Class-DispersionFullSchmatic}
   E \sim \sqrt{\sqrt{\lambda} (J_1+J_2)} \Big(1 + \frac{1}{\sqrt{\lambda}} \big(a_0^{(0)} (J_1+J_2)+ \frac{a_1^{(0)}}{J_1+J_2} + a^{(1)} \big) \\+ \frac{1}{\lambda} \big(b_0^{(0)} (J_1+J_2)^2 + \frac{b_1^{(0)}}{(J_1+J_2)^2} + b_0^{(1)} (J_1+J_2) + \frac{b_1^{(1)}}{J_1+J_2} + b^{(2)} \big) + \mathcal{O}({\lambda^{-\frac{3}{2}}})  \Big)\, .
\end{multline}
More precise formulas will be given in the following sections.
The energy above is the direct generalisation of the flat space energy, with the leading order term being the \textit{flat-space} string energy. The higher terms are curvature corrections to the flat space result.

Terms in the semiclassical expansion organise in a simple way. For instance, the leading terms at each order in the $1/\sqrt{\lambda}$ expansion are completely fixed by the \textit{classical} string solution. In the schematic example above, the four coefficients $a_0^{(0)},a_1^{(0)}$, $b_0^{(0)},b_1^{(0)}$ are determined by the classical solution. The subleading terms $a^{(1)}$ and $b_0^{(1)},b_1^{(1)}$ are determined by the 1-loop fluctuation around the classical solution, while the sub-subleading term $b^{(2)}$ is determined by the 2-loop fluctuation. Higher terms in the $1/\sqrt{\lambda}$ expansion follow the same pattern.

After finding the semiclassical expansion, the next step involves extrapolating the analysis to the regime of small charge, characterised by $J_i, {\bf S}_i = \mathcal{O}(1)$, or equivalently, where $\mathcal{J}_i, \mathcal{S}_i \ll 1$ and trying to fit the operator into the long multiplet of the superconformal primary operator we are interested in. 
As operators in the same long multiplet are related by supersymmetry, this fitting allows us to determine the spectrum of the superconformal primary. This extrapolation is non-rigorous but works well for the $\mathcal{N}=4$ case and we will assume it works for the $\mathcal{N}=2$ open string case as well. 

\subsection{Classical open string solutions} \label{sec:classicalOpenCond}

We first derive the leading terms at each order from a classical solution for open strings. The action, the equations of motion and the expression for the conserved charges of the open strings are identical to the closed string case, see appendix \ref{apd:classical} for details. The differences include:
\begin{itemize}
    \item The worldsheet parameter $\sigma$ spans $[0,\pi)$ for open strings, in contrast to the $[0,2\pi)$ range for closed strings.
    \item Open strings end on $D$-branes, where classically, fermions vanish, leaving only bosonic fields to contribute to the classical dynamics. These fields satisfy Dirichlet or Neumann boundary conditions at $\sigma =0$ or $\sigma =\pi$, instead of being periodic.\footnote{It is much more involved to study the allowed fermionic boundary conditions as one has to work in Green-Schwartz formalism to incorporate the RR-flux. See \cite{Mann:2006rh,Dekel:2011ja,Linardopoulos:2021rfq,Linardopoulos:2022wol} for some general discussions in the absence of the orientifold. }
    \item In our case, involving a configuration of $N$ $D3$ branes, four $D7$ branes, and one $O7$ plane all positioned together, the introduction of $D7$ branes breaks the $SO(6)$ $R$ symmetry down to its subgroup $SO(4) \times SO(2)$ which is isomorphic to $SU(2)_L \times SU(2)_R \times U(1)_R$, resulting in a partial breaking of the $\mathcal{N}=4$ supersymmetry to $\mathcal{N}=2$.
    Here, the $SO(4)$ symmetry represents the transverse rotation along the $D7$ brane and the $U(1)_R$ corresponds to the longitudinal rotations of the $D7$ brane. Notably, we assume the orientifold plane retains the $SO(4)$ rotational symmetry among $J_1, J_2$ directions.\footnote{$J_1, J_2$ directions still satisfy the Neumann boundary condition as they lie on the $D7$ brane, while the $J_3$ direction satisfies the Dirichlet boundary condition because it is the transverse direction of the $D7$ brane.}
    In this case, the Cartans of the $\mathcal{N}=2$ theory are related to the $\mathcal{N}=4$ theory's $J_1, J_2, J_3$ as follows\footnote{The simplest way to confirm the numerical prefactor is to compare the Dynkin label of the $\mathcal{N}=2$ supercharge in \cite{Cordova:2016emh} with the Dynkin labels of the $\mathcal{N}=4$ supercharge in \cite{Dolan:2002zh}.}
    \beq \label{eqn-N4toN2}
    J_{SU(2)_R} = J_1 + J_2, \quad J_{SU(2)_L} = J_1 - J_2, \quad J_{U(1)} = 2J_3\, .
    \eeq
    \item Our theory is defined on an orientifold and geometrically the orientifold $O7$ plane lies on top of the $D7$ plane. Therefore, we have to identify the transverse directions of $D7$ brane by a $\mathbb{Z}_2$ orbifold action, and we only keep the states that are invariant under the worldsheet parity plus the $\mathbb{Z}_2$ orbifold action.
\end{itemize}
The last point needs clarification.

\paragraph{The orientifold action.}

We begin with the orientifold action in our geometric setup. Here, the four probe $D7$ branes wrap an $S^3$ inside $S^5$ of the curved $AdS_5 \times S^5$ spacetime. In terms of global coordinates \eqref{eqn-GlobalCoord}, this necessitates to set two out of the six $X_i$'s to be zero. The natural choices are given by setting $\cos \gamma =0$, $\sin \psi =0$ or $\cos \psi =0$. The orientifold action, denoted by $\Omega$, is generated by composing three different operations:
\beq \label{eqn-Class-OrientifoldCond}
\Omega = (-)^{\mathcal{F}}\, \tilde{\Omega}\,  \mathcal{I}\, .
\eeq
Here, the operator $(-)^{\mathcal{F}}$ counts the fermion numbers; the second operator $\tilde{\Omega}$ imposes worldsheet parity as follows
\beq
\tilde{\Omega}: \quad (\tau, \sigma) \rightarrow (\tau, \pi - \sigma)\, .
\eeq
The last operator, $\mathcal{I}$, imposes the $\mathbb{Z}_2$ orbifold action in the transverse direction of the $D7$ plane. 

The explicit action of $\mathcal{I}$ varies depending on the embedding of the branes within the coordinate system, and the results are summarised in the following table.
\begin{table}[!ht]
    \centering
    \begin{tabular}{c|c|c} \hline
       Condition  & Angle Value & Orbifold Action \\ \hline
     $\cos \gamma =0$    & $\gamma = \pi/2$  & $\varphi_1 \rightarrow \varphi_1 +\pi$ \\ \hline 
     $\cos \psi =0$    & $\psi = \pi/2$  & $\varphi_2 \rightarrow \varphi_2 +\pi$ \\ \hline   
     $\sin \psi =0$    & $\psi = 0$  & $\varphi_3 \rightarrow \varphi_3 +\pi$ \\ \hline   
    \end{tabular}
    \label{tab:orbifold_actions}
\end{table}

The open string state also carries the Chan-Paton factor $\beta^{I_1I_2}$, where $I_1,I_2$ are the usual adjoint indices. The orientifold action for any product of orthogonal groups (which includes all the $G_F$ we consider) simply permutes the indices \cite{Gimon:1996rq}
\beq
\beta^{I_1I_2} \to  \beta^{I_2I_1}\, .
\eeq
Thus, for symmetric irreps ${\bf1}$ and ${\bf sym}$ we get a plus sign, while for the antisymmetric irrep $\bf adj$ we get a minus sign. The final requirement for the invariance of the classical solution is for the combined solution with the Chan-Paton factor to be invariant under the orientifold action.

\paragraph{The constraint on winding numbers.}

Implementing the orientifold condition \eqref{eqn-Class-OrientifoldCond} on the classical solution is a well-defined procedure. However, there is a subtle point worth mentioning during the final step of extrapolating to small quantum numbers. To ensure the invariance of the classical solution under the orientifold action, it is necessary to demand \textit{each component} of the solution to be invariant. This requirement imposes specific constraints on the winding number $m$ of the solution. In most cases, these constraints limit $m$ to even integers, denoted as $m \in 2\mathbb{Z}$.

Considering the oscillator representations of the state reveals that the previously mentioned requirement might be too stringent. To elaborate this point, let us begin with a flat space string example, assuming the existence of two independent oscillators $a_{n_1}^\dagger$ and $ a_{n_2}^\dagger$, each at a different level $n_j$, to construct the state. The state characterised by the quantum numbers $(J, S)$ and Chan-Paton factor $\beta^{I_1I_2}$ is constructed as
\beq
\beta^{I_1I_2} |J, {\bf S} \rangle \equiv\beta^{I_1I_2}(a_{n_1}^\dagger)^{J} (a_{n_2}^\dagger)^{\bf S}  |0\rangle\,,
\eeq
where $|0\rangle$ represents the vacuum state. The orientifold action $\Omega$ modifies the sign of each oscillator based on their level, as described by
\beq
\Omega a_{n_j}^\dagger\Omega^{-1} = (-)^{n_j} a_{n_j}^\dagger, \quad j = 1,2, \quad \text{and} \quad \Omega |0\rangle = |0\rangle, 
\eeq
resulting in the transformation of the state under $\Omega$ to:
\beq
\Omega\beta^{I_1I_2} |J, {\bf S} \rangle = (-)^{n_1 J + n_2 {\bf S} } \beta^{I_2I_1}|J, {\bf S} \rangle \, .
\eeq
Thus, the criterion for the invariance of the state under the orientifold action is for the \textit{total} quantum number $n_1 J + n_2 {\bf S} $ to be even for symmetric irreps, or odd for the antisymmetric irrep. 

For classical solutions, we impose the orientifold condition on each $X$ separately, which would imply that each component is even, and thus requires $n_1 J$ and $n_2 {\bf S}$ to be \textit{separately} even and only for symmetric irreps to exist.
This requirement can be intuitively understood through the nature of the semiclassical expansion: in the semiclassical regime, $J$ and ${\bf S}$ are treated as large \textit{real} parameters that are not necessarily integers. Therefore, to maintain invariance under the orientifold action without the integrality of the $J$ and $S$, we must ensure that $n_1$ and $n_2$ are even.

For strings propagating in $AdS_5 \times S^5$, the same issue persists. Imposing the orientifold condition directly in the large $J_i,{\bf S}_i$ semiclassical limit seems too strong a constraint, as it requires even winding number and restricts to symmetric irreps. Instead, we will only require that the orientifold condition be satisfied after extrapolating $J_i,{\bf S}_i$ to the finite integers we are interested in, and we will interpret the orientifold constraint in the oscillator sense considered above, even if it is not yet precisely understood on $AdS_5 \times S^5$.
 
\subsection{Open string solutions from $\mathcal{N}=4$ closed string solutions}

Guided by the conditions outlined previously, we proceed to solve the string equations of motion (EOMs). Instead of tackling the EOMs directly, we find inspiration in the solutions already obtained for closed strings. It is noteworthy that both open and closed strings satisfy the same EOMs and they are connected through the ``doubling trick''. This approach allows us to use the information of the existing closed string solutions to search for open string solutions \cite{Stefanski:2003qr,Chen:2004mu,Chen:2004yf,Okamura:2005cj}.

More specifically, the doubling trick allows for generating the corresponding closed string solution from an open string solution by gluing two identical copies of the open string solution. It is important to note, however, that the converse of this procedure is not automatically true. While one can limit the domain of $\sigma$ from $[0,2\pi)$ to $[0,\pi)$ to generate a function that solves the open string EOMs, one still needs to ensure that this adjusted solution obeys all previously mentioned boundary and orientifold conditions.

The known classical closed $\mathcal{N}=4$ string solutions are of the following types:
\begin{equation}
\centering
 \begin{tabular}{c|c|c} \hline
   \text{Type}  & \makecell{$\mathcal{N}=4$  Quantum Number \\ $(J_1,J_2,J_3)$} & \makecell{$\mathcal{N}=2$  Quantum Number \\ $(J_{SU(2)_R},J_{SU(2)_L},J_{U(1)_R})$} \\ 
   \hline
    Circular $S^5$ & $(0,0|J_1,J_3, J_1)$ & $(0,0|J_1+J_3,J_1-J_3,2J_1)$ \\ \hline
    Circular $AdS_5$ & $({\bf S},{\bf S}|J,0, 0)$ & $({\bf S},{\bf S}|J,J,0)$ \\ \hline
    Circular Mixed & $({\bf S},0|{\bf S},0, J)$ & $({\bf S},0|{\bf S},{\bf S},2J)$ \\ \hline
    Folded Mixed & $({\bf S},0|J,0,0)$ & $({\bf S},0|J,J, 0)$ \\ \hline
    Folded $S^5$ & $(0,0|J_1,J_2,0)$ & $(0,0|J_1+J_2, J_1-J_2, 0)$ \\ \hline
    Glued Folded & $({\bf S},0|J_1,J_2,0)$ & $({\bf S},0|J_1+J_2, J_1-J_2, 0)$ \\ \hline
\end{tabular}   
\end{equation}
The glued folded solution generalises the two different types of folded solution and interpolates between them smoothly. Therefore we only need to consider the glued folded solution.

Using the mapping in \eqref{eqn-N4toN2}, we deduce the quantum numbers for the $\mathcal{N}=2$ solution, if this restriction is legitimate. Here the identification of the $\mathcal{N}=4$ directions with $J_1, J_2$ direction is achieved by examining the boundary conditions of the bosonic field, where we recall that $J_1$ and $J_2$ correspond to the quantum numbers of the two Neumann directions.

As discussed in the preceding section, our goal is to identify states that are a singlet under the $SU(2)_L$ symmetry of the $\mathcal{N}=2$ theory. This objective imposes restrictions on the allowed quantum numbers of classical solutions. Further analysis of the boundary conditions of the bosonic fields leads us to conclude that merely restricting some closed solutions does not yield valid open solutions that satisfy the appropriate boundary conditions. Lastly, we want a solution that will apply for every Lorenz spin $\ell$. The relevant solution for us, is the glued folded solution, with quantum numbers
\begin{equation}
\centering
 \begin{tabular}{c|c|c} \hline
   \text{Type}  & $\mathcal{N}=4$  Quantum Number & $\mathcal{N}=2$ Quantum Number \\ \hline
    Glued Folded & $({\bf S},0|J,J,0)$ & $({\bf S},0|2J, 0, 0)$ \\ \hline
\end{tabular}   
\end{equation}

Now we want to analyse whether this glued folded solution can be fitted into the SUSY multiplet of the superconformal primaries that appear in the AdS Veneziano amplitude.  Recall from the previous section that the superconformal primaries identified are neutral under the $R$-symmetry and carry identical Lorentz spin $\ell=\mathtt{j} = \bar{\mathtt{j}}$. Using the dictionary \eqref{eqn-DynkinDict}, they correspond to states of the quantum number $(\ell,0|0,0,0)$. 
The analysis reveals that, with appropriately selected values of $J$, the folded string solution indeed qualifies as a superconformal descendant of this primary state. A detailed examination of the long multiplets table in section 4.6 of \cite{Cordova:2016emh} shows that the glued folded string solution matches a state in the middle column of the long multiplet, characterised by having an $R$-charge two units higher than the primary state and retaining the same spin for $J=1$.\footnote{Additionally, another possible fitting is the one where for the primary spin $[\mathtt{j}, \bar{\mathtt{j}}]^{R=0}$, the glued folded string aligns with a state exhibiting spins $[\mathtt{j}\pm 1, \bar{\mathtt{j}} \pm 1]^{R=2}$. However, only the current identification fits the solution from the previous sections.}

\subsection{The classical glued folded string solution}

The classical glued folded string solution takes the following form 
\beq
\theta =0, \quad \rho = \rho(\sigma), \quad \phi_3 = k \tau, \quad \gamma = \frac{\pi}{2}, \quad \psi = \psi(\sigma), \quad \varphi_1 =\omega_1 \tau, \quad \varphi_2 = \omega_2 \tau.
\eeq
The equations of motion and Virasoro constraints can be solved, and details will be given in appendix \ref{apd:classical}. The resulting classical energy reads
\bea
        E_{\textrm{cl}} &=  \sqrt{m\sqrt{\lambda} ({\bf S}+J_1)} \bigg(1 + \frac{1}{\sqrt{\lambda} m} \left(\frac{3{\bf S}}{4} + \frac{J_1}{4} +\frac{J^2_2}{2({\bf S}+J_1)} \right) \\
        &+ \frac{1}{\lambda m^2} \left(\frac{3 J_1^2}{32}-\frac{J_2^4}{8 (J_1+{\bf S})^2}+\frac{J_1 J_2^2}{4 (J_1+{\bf S})}-\frac{13 J_1 {\bf S}}{16}+\frac{5 J_2^2}{8}-\frac{21 {\bf S}^2}{32}\right)  + \mathcal{O}(1/\lambda^\frac{3}{2}) \bigg)\, ,
\eea{eqn-E-class-GluedFolded}
where $m \in \mathbb{Z}$ is the winding number. The requirements for achieving an $SU(2)_L$ singlet state within the desired supermultiplet set both $J_1$ and $J_2$ equal to one. The lowest energy state is then given by setting $m=1$. As discussed above, we impose the orientifold condition in the oscillator sense after extrapolating $J_i=1$ and ${\bf S}=\ell$, which means that for even $\ell$ we have symmetric irreps $\bf1$ and $\bf sym$, while for odd $\ell$ we have the antisymmetric irrep $\bf adj$, as expected. Based on the general discussion around \eqref{eqn-Class-DispersionFullSchmatic}, we then find that the full quantum energy up to this order can be expressed as follows
\bea
        E ={}&  \sqrt{\sqrt{\lambda} (\ell+1)} \bigg(1 + \frac{1}{\sqrt{\lambda} } \left(\frac{3\ell}{4} + \frac{1}{4} +\frac{1}{2(\ell+1)} + a^{(1)} \right) \\
        &+ \frac{1}{\lambda } \left(-\frac{21 \ell^2}{32}-\frac{1}{8 (1+\ell)^2}+ b_0^{(1)}  \ell+  \frac{b_1^{(1)}}{1+\ell}+ b^{(2)} \right)  + \mathcal{O}({\lambda^{-\frac{3}{2}}}) \bigg)\,.
\eea{Equant}
We find that the $O(1/\sqrt{\lambda})$ terms match the spin-dependent parts of our previous result \eqref{tau2}, where recall that the superprimary spin $\ell$ is related to $\delta$ as $\ell = \delta -1$. In particular, we find that both the odd spin $\bf adj$ and even spin ${\bf1}$ and $\bf{sym}$ irreps are given by the same formula. The $1/\lambda$ term is a prediction for the next order.

\section{Conclusion}
\label{conc}

In this paper, we computed the first curvature correction to the AdS Veneziano amplitude for general flavour group $G_F$ by combining a dispersion relation with an ansatz for the worldsheet integral for this amplitude. We checked our solution in three independent ways. Firstly, we showed that the exponent of our solution in the high energy limit is half of the corresponding exponent for closed strings as expected \cite{Gross:1989ge}. Secondly, we showed that the $1/\sqrt{\lambda}$ correction to the dimensions of massive string operators matches an independent semiclassical calculation for open strings on $AdS_5\times S^5/\mathbb{Z}_2$.
Thirdly, the low energy expansion of our result is consistent with previous results from localisation  obtained in \cite{Behan:2023fqq} for the case $G_F=SO(8)$.
We also combined our solution with the constraints of \cite{Behan:2023fqq} to fix the $\lambda^{-2}$ correction at finite $R$, which corresponds to the tree level unprotected $D^4F^4$ correction to the super-Yang-Mills action on $AdS_5\times S^3$.

Our method of combining dispersion relations with a worldsheet integral ansatz can also be used to constrain the AdS Veneziano amplitude to higher orders in $1/R$, as was the case with the AdS Virasoro-Shapiro amplitude. In the latter case, integrability results for massive string operators were then sufficient to fix the next order in the curvature expansion \cite{Alday:2023mvu}, and likely higher orders too. In our open string case, integrability has not yet been worked out for the classical worldsheet theory, which is why we instead had to use a semiclassical expansion to compute the spin dependent terms of the first $1/\sqrt{\lambda}$ correction to the massive string operators.\footnote{To fix the complete $1/\sqrt{\lambda}$ correction would require a 1-loop correction to the classical string solution we considered, which seems challenging due to the subtlety of imposing the orientifold constraint.} The semiclassical expansion involves a non-rigorous extrapolation to finite quantum numbers, which is especially subtle when implementing the orientifold constraint. A rigorous integrability analysis would be useful both to check this semiclassical expansion, as well as allow us to compute higher orders in the curvature expansion of the AdS Veneziano amplitude. We are looking into this and hope to report back soon.

Now that both the AdS Veneziano and Virasoro-Shapiro amplitudes are available, at least to the first couple orders in a small curvature expansion, it would be interesting to compare them. In flat space, the famous KLT relation \cite{KAWAI19861} shows that the Virasoro-Shapiro amplitude is the square of the Veneziano amplitude, up to a phase factor. This relation can also be understood by deforming the contour of the two dimensional worldsheet integral for the Virasoro-Shapiro amplitude into two copies of the one dimensional worldsheet integral of the Veneziano amplitude. In Appendix \ref{facApp}, we show some first steps to generalising this for the AdS amplitudes, but did not find any obvious relation.

One of the checks on our curvature correction to the AdS Veneziano amplitude was the low energy expansion for the $G_F=SO(8)$ theory, which was computed by combining localisation with analytic bootstrap in \cite{Behan:2023fqq}. It would be nice to generalise this calculation to the other two theories with $U(4)$ and $SO(4)\times SO(4)$ flavour groups. The challenge in these cases is that the matrix model for the localised mass deformed partition function is no longer a single trace deformation of a free matrix model, which facilitated the calculation of \cite{Behan:2023fqq,Beccaria:2021ism,Beccaria:2022kxy}. 

Finally, the general method of combining dispersion relations with an ansatz for the worldsheet integral should work for other cases of AdS/CFT with a weakly coupled string theory limit. One such example is type IIA string theory on $AdS_4\times \mathbb{CP}^3$, which is dual to ABJM theory with $U(N)_k\times U(N)_{-k}$ gauge group in the large $N,k$ limit \cite{Aharony:2008ug}. Both integrability \cite{Cavaglia:2014exa} and the superblock expansion \cite{Binder:2019mpb,Binder:2020ckj} of the stress tensor multiplet is known for this theory, which should allow us to compute curvature corrections to the AdS Virasoro-Shapiro amplitude to all orders in $\lambda\equiv N/k$.\footnote{The first couple $1/\lambda$ corrections to this holographic correlator was already studied at finite $R$ in \cite{Binder:2019mpb}.}

\section*{Acknowledgements} 

We thank Connor Behan, Joao Silva, Bogdan Stefanski and Arkady Tseytlin for useful conversations. The work of LFA and TH is supported by the European Research Council (ERC) under the European Union's Horizon 2020 research and innovation programme (grant agreement No 787185). LFA is also supported in part by the STFC grant ST/T000864/1. 
TH is also supported by the STFC grant ST/X000591/1.
SMC and DlZ are supported by the Royal Society under the grant URF\textbackslash R1\textbackslash 221310. 

\appendix

\section{Mack polynomials}\label{app:mack}

Our definitions for the Mack polynomials are
\begin{align}
\mathcal{Q}_{\ell, m}^{\tau, d}(t) = K(\tau,\ell,m,3) Q_{\ell, m}^{\tau, d}(t) \,, 
\label{curlyQ}
\end{align}
with
\beq
K(\tau, \ell, m,\Delta)  = - \frac{2 (\ell+\tau -1)_\ell \Gamma (2 \ell+\tau )}{ 2^\ell \Gamma \left(\ell+\frac{\tau}{2} )\right)^4 \Gamma (m+1) \Gamma \left(\Delta -\frac{\tau }{2}-m\right)^2 \left(\ell+\tau-\frac{d}{2} +1\right)_m}.
\eeq
Note that the value for $\Delta$ changes for different reduced Mellin amplitudes. For the full Mellin amplitude of four identical scalars $\phi$ it is $\Delta=\Delta_\phi$, while for the reduced Mellin amplitude for the $\langle 2222 \rangle$ correlator in $\mathcal{N} = 4$ SYM, we have to set $\Delta=4$.
$Q_{\ell, m}^{\tau, d}(t)$ is called a Mack polynomial in the literature \cite{Mack:2009mi}. We found the following representation useful \cite{Dey:2017fab}
\begin{align}
&Q_{\ell, m}^{\tau, d}(t) = (-1)^{\ell} 4^{\ell} \sum _{n_1=0}^{\ell} \sum_{m_1=0}^{\ell-n_1} (-m)_{m_1} \left(m+\frac{t}{2}+\frac{\tau }{2}\right)_{n_1} \tilde{\mu}(\ell,m_1,n_1,\tau ,d), \\
&\tilde{\mu}(\ell,m, n,\tau ,d) \equiv \frac{2^{-\ell} \Gamma (\ell+1) (-1)^{m+n} \left(\ell-m+\frac{\tau }{2}\right)_m \left(n+\frac{\tau }{2}\right)_{\ell-n} }{\Gamma (m+1) \Gamma (n+1) \Gamma (\ell-m-n+1)}  \\
& \times \left(\frac{d}{2}+\ell-1\right)_{-m} (2 \ell+\tau -1)_{n-\ell} \left(m+n+\frac{\tau }{2}\right)_{\ell-m-n} \nonumber \\
& \times \, _4F_3\left(-m,-\frac{d}{2}+\frac{\tau }{2}+1,-\frac{d}{2}+\frac{\tau }{2}+1, \ell+n+\tau -1;\ell-m+\frac{\tau }{2},n+\frac{\tau }{2},-d+\tau +2;1\right) . \nonumber 
\end{align}

\section{Dispersive sum rules}
\label{sec:dsr}

In this section we use a crossing-symmetric dispersion relation to directly compute the Wilson coefficients in the expansion \eqref{wilson_expansion} in terms of OPE data.

\subsection{Crossing-symmetric dispersion relation}

A crossing-symmetric dispersion relation for Mellin amplitudes with the symmetry
\beq
M(s,t) = M(t,s)\,,
\eeq
was first derived in \cite{Fardelli:2023fyq}, and we use essentially the same relation.
We express the Mellin amplitude in terms of the crossing-symmetric variables $u$ and $r= \frac{\sigma_2}{\sigma_1}$
\beq
\tilde{M}(u,r)=M (s'(u,r),t'(u,r))\,,
\eeq
by using
\beq
s'(u,r) = \frac{1}{2} \left(-u + \sqrt{u(u-4 r)}\right)\,, \qquad
t'(u,r) = \frac{1}{2} \left(-u - \sqrt{u(u-4 r)}\right)\,,
\label{s23_to_s1r}
\eeq
which solves $r = \sigma_2 / \sigma_1$ and satisfies $u + s'(u,r) + t'(u,r) = 0$.
At large $u$ we have either
\beq
s'(u,r) = - r + O\left( \frac{1}{u} \right)\,, \qquad
t'(u,r) = -u + r + O\left( \frac{1}{u} \right)\,,
\eeq
or
\beq
s'(u,r) = -u + r  + O\left( \frac{1}{u} \right)\,, \qquad
t'(u,r) = - r + O\left( \frac{1}{u} \right)\,,
\eeq
so the bound on chaos \eqref{boc} with fixed $s$ or $t$
translates to the following bound for fixed $r$
\beq
\tilde{M}(u, r) = o(u^{-1})\,, \text{ for } |u|\to \infty \text{ with } \text{Re}(r)> 0\,.
\label{boc_fixed_r}
\eeq
We can now use the bound \eqref{boc_fixed_r} to write a dispersion relation starting with
\beq
\tilde{M}(u, r) = \oint_{u} \frac{du'}{2 \pi i}  \frac{\tilde{M}(u', r )}{(u'-u)}\,.
\label{dr_start}
\eeq
In terms of $u, s, t$ and $\tau_{m} = \tau + 2m-2$ the Mellin amplitude has poles
\bea
M(s,t) &\approx C_{\tau, \ell}^{2} \frac{\mathcal{Q}_{\ell, m}^{\tau +2, 4}(u-2)}{s-\tau_{m}} \,, \\
M(s,t) &\approx C_{\tau, \ell}^{2} \frac{\mathcal{Q}_{\ell, m}^{\tau +2, 4}(u-2)}{t-\tau_{m}} \,.
\eea{mellin_poles}
We assume that poles in the $u$-channel are absent altogether, as is the case in flat space (for colour-ordered amplitudes).
$\tilde{M}(u,r)$ is a meromorphic function in $u$ due to the symmetry $M(s,t)=M(t,s)$.
The poles lie at $u=-\frac{\tau_m^2}{\tau_m+r}$, which corresponds to $s'(u,r)= \tau_m$ or $t'(u,r)= \tau_m$, depending on the values of $r$ and $\tau_m$.
In this way we finally get the dispersion relation
\bea
\tilde{M}(u, r) ={}& - \sum_{\tau,\ell,m} C_{\tau, \ell}^{2} 
\frac{\tau _{m} \left(\tau _{m}+2 r\right)}{\tau _{m}+r}
\frac{\mathcal{Q}_{\ell, m}^{\tau +2, 4}(-\frac{\tau_{m}^2}{\tau_{m} +r}-2)}{\tau_{m}^2 + (\tau_{m} +r) u} \,.
\eea{dispersion}
Note that all the Mack polynomials depend only on $r$, so if we expand
\beq
\tilde{M}(u, r) = \sum_{a,b} \alpha_{a,b} \sigma_1^a \sigma_2^b  = \sum_{a,b} \alpha_{a-b,b} \sigma_1^a r^b\,,
\eeq
low values of $b$ correspond to simple sum rules for $\alpha_{a,b}$.
In order to do the Taylor expansions in $r$ of the Mack polynomials in \eqref{dispersion} we apply the chain rule to a generic function $f(s)$
\bea
\partial_r^n f\left(-\tfrac{\tau_{m}^2}{\tau_{m} +r}\right) \big|_{r=0} &= 
\sum\limits_{q=1}^n 
\frac{(-1)^{n-q}  (n-q+1)_{q-1} (q+1)_{n-q} }{\Gamma (q)\tau _m^{n-q}}
\partial_{s}^q f(s)\big|_{s=-\tau_m} \,, \ && n>0\,.
\eea{taylor_derivatives}
We further use $\mathcal{Q}_{\ell, m}^{\tau, 4}(s) = \mathcal{Q}_{\ell, m}^{\tau, 4}(-s-2-\tau-2m)$ to write
\beq
\partial_{s}^q \mathcal{Q}_{\ell, m}^{\tau +2, 4}(s-2)\big|_{s=-\tau_m}
= (-1)^q \partial_{s}^q \mathcal{Q}_{\ell, m}^{\tau +2, 4}(s-2)\big|_{s=0}\,.
\eeq
By expanding also the other factors in $u$ and $r$, we find for the Wilson coefficients
\bea
\alpha_{a-b,b} ={}& \sum_{\tau,\ell,m} C_{\tau, \ell}^{2} \sum_{q=0}^b \mathfrak{U}_{a,b,q}(\tau_m) \partial_{s}^q \mathcal{Q}_{\ell, m}^{\tau +2, 4}(s-2)\big|_{s=0}\,,
\eea{alpha}
with
\bea
\mathfrak{U}_{a,b,0}(\tau_m) &= \frac{(a+b) (1-a)_{b-1} }{(-1)^{b} \Gamma (b+1)\tau_m^{a+b+1}}\,,\\
\mathfrak{U}_{a,b,q>0} (\tau_m) &=
\frac{\pi   \Gamma (a)   {}_4\tilde{F}_3(1,1,1-b,-a-b+2;-a-b+1,a-b+2,2-q;1)}{(-1)^{b}\sin (\pi  a) \Gamma (b) \Gamma
   (q) \Gamma (q+1) \Gamma (a+b-1) \tau_m^{a+b+1-q}}\,,
\eea{Ufrak}
where ${}_4\tilde{F}_3$ is the regularised hypergeometric function.

\subsection{$1/\lambda$ expansion}

The first dispersive sum rule that we obtain by expanding \eqref{alpha} with the OPE data \eqref{ope_data_expansion} (but leaving the leading twist $\tau_0$ unfixed) is
\beq
\alpha^{(0)}_{a,0} = \sum\limits_{\cO_{\tau,\ell}} \frac{f_0}{(\tau_0^2)^{a+2}} \,.
\eeq
By comparing this with \eqref{alpha0_from_flat} in the limit of large $a$ we see that
\beq
\tau_{0}(r) = \sqrt{\delta}\,, \qquad \delta \in \mathbb{N}^+\,.
\eeq
Using this, we get the first layer of sum rules in terms of $\delta$
\begin{equation}
\alpha^{(0)}_{a,b} = \sum\limits_{\delta=1}^\infty \sum\limits_{q=0}^b  \frac{c_{a,b,q}}{\delta^{2+a+2b}} F_q^{(0)}(\delta)\,,
\label{alpha0}
\end{equation}
with
\bea
c_{a,b,0} ={}& \frac{ (a+2 b) (-a-b+1)_{b-1}}{(-1)^{b+1}\Gamma (b+1)}\,,\\
c_{a,b,q>0} ={}& \frac{\pi  \Gamma (a+b) \, _4\tilde{F}_3(1,1,1-b,-a-2 b+2;-a-2
   b+1,a+2,2-q;1)}{(-1)^{b+1} \sin (\pi  (a+b)) \Gamma (b) \Gamma (q) \Gamma (a+2 b-1)}\,,
\eea{c0_def}
and
\beq
F_q^{(0)}(\delta) = \frac{4^q}{\Gamma(2q+2)} \sum_{\ell=0}^{\delta-1}  (\ell-q+1)_q (\ell+2)_q  \langle f_0 \rangle_{\de,\ell}\,.
\label{F0_def}
\eeq
By comparing the sum rule with \eqref{alpha0_from_flat}, we find the solution, which is simply an Euler-Zagier sum with $q$ 1's
\beq
F_q^{(0)}(\delta) = \delta^q Z_{\underbrace{1,\ldots,1}_{q}} (\delta-1)\,.
\label{F0_solution}
\eeq
The generating series for this solution is
\beq
\sum\limits_{q=0}^\infty F_q^{(0)}(\delta) \left( \frac{z}{\delta} \right)^q = \binom{z+\delta-1}{\delta-1}\,.
\label{F0_generating_series}
\eeq

At the next order the requirement that \eqref{wilson_expansion} is an expansion in $1/\sqrt{\lambda}$ leads to a sum rule for vanishing Wilson coefficients
\begin{equation}
0 = \sum\limits_{\delta=1}^\infty \sum\limits_{q=0}^b  \frac{c_{a,b,q}}{\delta^{\frac52+a+2b}} \left( F_q^{(1)}(\delta) - (2+ a+2 b) T_q^{(1)}(\delta) \right)\,,
\label{zero_sum_rule}
\end{equation}
with
\bea
T_q^{(1)}(\delta) ={}& \frac{4^q}{\Gamma(2q+2)} \sum_{\ell=0}^{\delta-1}  (\ell-q+1)_q (\ell+2)_q
\langle f_0 \rangle_{\de,\ell} 2(\tau_1(\delta,\ell) + \ell) \,,\\
F_q^{(1)}(\delta) ={}& \frac{4^q}{\Gamma(2q+2)} \sum_{\ell=0}^{\delta-1}  (\ell-q+1)_q (\ell+2)_q
\left( \sqrt{\delta }  \langle f_1\rangle_{\de,\ell} -\langle f_0\rangle_{\de,\ell} ( 4 \ell- \tfrac{1}{2} ) \right)\,.
\eea{T1F1_def}
This has the solution
\begin{align}
\tau_1(\delta, \ell)  = - \ell\,, \qquad \langle f_1 \rangle_{\de,\ell} = \langle f_0\rangle_{\de,\ell} \frac{4 \ell-\frac12}{\sqrt{\delta }}\,.
\end{align}
The next dispersive sum rule is
\begin{equation}
\alpha^{(1)}_{a,b}= \sum\limits_{\delta=1}^\infty \sum\limits_{q=0}^b  \frac{c_{a,b,q}}{\delta^{3+a+2b}} \left( F_q^{(2)}(\delta) - (2+ a+2 b) T_q^{(2)}(\delta)
+ p^{(2,0)}_{a,b,q} F_q^{(0)}(\delta) + p^{(2,1)}_{a,b,q} F_{q+1}^{(0)}(\delta) \right)
\,,
\label{alpha1_sum_rule}
\end{equation}
with
\bea
T_q^{(2)}(\delta) ={}& \frac{4^q}{\Gamma(2q+2)} \sum_{\ell=0}^{\delta-1}  (\ell-q+1)_q (\ell+2)_q
2 \sqrt{\delta} \langle f_0 \tau_2\rangle_{\de,\ell}\,,\\
F_q^{(2)}(\delta) ={}& \frac{4^q}{\Gamma(2q+2)} \sum_{\ell=0}^{\delta-1}  (\ell-q+1)_q (\ell+2)_q
\left( \delta \langle f_2\rangle_{\de,\ell} + 22 \ell \langle f_0\rangle_{\de,\ell} \right)\,,
\eea{T2F2_def}
and
\begin{align}
p^{(2,0)}_{a,b,q} ={}&
-q^2 (a+2 (b+7))+2 q (a+2 b) (a+2 b+6)-\frac{9 q}{2}-8 (a+2) b^2-4 a (a+4) b\nonumber\\
&-\frac{2}{3} a (a+1) (a+5)-\frac{16 b^3}{3}-\frac{20 b}{3}+\frac{59}{8}
\,,\label{p_def}\\
p^{(2,1)}_{a,b,q} ={}&
\frac{1}{4} (q+1) \left(4 a^2+2 a (8 b-2 q+11)+16 b^2-8 (b+7) q+44 b-37\right)\,.\nonumber
\end{align}

\subsection{Summing the low energy expansion}
\label{app:summing}

Let us also sum the low energy expansion, to get a new representation for the Veneziano amplitude. The main task is to derive a generating series for the coefficients $c_{a,b,q}$.
We start by noting that
\beq
c_{a,b,q} = - \sum\limits_{k=0}^{b-1} (-1)^k c_{a+1+k,b-1-k,q-1}\,,
\eeq
which lets us express these coefficients in terms of $c_{a,b,0}$
\beq
c_{a,b,q} = (-1)^q \sum\limits_{k_1,\ldots,k_q=0}^\infty (-1)^k c_{a+q+k,b-q-k,0}\,, \qquad k = k_1 + \ldots + k_q\,.
\eeq
Hence we can first do the sum
\beq
\sum\limits_{a,b=0}^{\infty} c_{a+q+k,b-q-k,0} x^a y^b = \frac{1+y}{1-x-y} \left( \frac{y}{1-y} \right)^{q+k}\,,
\eeq
which implies
\beq
\sum\limits_{a,b=0}^{\infty} c_{a,b,q} x^a y^b = \frac{1+y}{1-x-y} (-y)^q\,.
\label{c_generating_series}
\eeq
Together with the sum rule \eqref{alpha0} and the generating series \eqref{F0_generating_series} we can use this to show
\beq
\sum\limits_{a,b=0}^{\infty} \hat\sigma_1^a \hat\sigma_2^b \alpha^{(0)}_{a,b}
= \sum\limits_{\delta=1}^\infty
\frac{1}{\delta^2} \frac{1+y}{1-x-y} \binom{\delta-\delta y - 1}{\delta-1}\,,
\eeq
where
\beq
x= \frac{\hat\sigma_1}{\delta}\,, \qquad
y= \frac{\hat\sigma_2}{\delta^2}\,.
\eeq
Similarly one can use \eqref{c_generating_series} together with \eqref{alpha1_sum_rule}
to compute
\beq
A^{(1)}(S,T) = \sum\limits_{a,b=0}^{\infty} \hat\sigma_1^a \hat\sigma_2^b \alpha^{(1)}_{a,b} = \sum\limits_{i=1}^4 \frac{R^{(1)}_i(T,\delta)}{(S-\delta)^i} + O((S-\delta)^0)\,,
\eeq
with the numerators (where $C^{(\alpha)}_\ell(x)$ are Gegenbauer polynomials)
\begin{align}
{}&R^{(1)}_4(T,\delta) = - \frac{4 \delta}{\Gamma(\delta)} (T+1)_{\delta-1}\,,\nonumber\\
{}&R^{(1)}_3(T,\delta) = \frac{1}{2} \partial_T R_4(T,\delta) \,,\nonumber\\
{}&R^{(1)}_2(T,\delta) = \sum\limits_{\ell=0}^{\delta-1} 
\langle f_0  \rangle_{\delta ,\ell} \left(\frac{(\delta +2 T) C^{(1)}_{\ell+1}\left(\frac{2 T}{\delta }+1\right)}{2 T^2+2 \delta  T}-\frac{\left(2 \delta ^2-2 \ell^2 T (\delta +T)+\delta ^2 \ell\right)
   C^{(1)}_\ell\left(\frac{2 T}{\delta }+1\right)}{2 \delta  (\ell+1) T (\delta +T)}\right)\nonumber\\
&-\sum\limits_{\ell=0}^{\delta-1} \frac{2 \langle f_0 \tau_2 \rangle_{\delta ,\ell}C^{(1)}_\ell\left(\frac{2 T}{\delta }+1\right)}{\sqrt{\delta }
   (\ell+1)}\,,\label{R}\\
{}&R^{(1)}_1(T,\delta) = \sum\limits_{\ell=0}^{\delta-1} 
\langle f_0 \rangle_{\delta ,\ell} \bigg[  
\Big(\left(8 \ell^3+216 \ell^2-44 \ell-9\right) T^4
+3 \delta  \left(4 \ell^3+208 \ell^2-44 \ell-9\right) T^3\nonumber\\
&+\delta ^2 \left(4 \ell^3+636 \ell^2-55\ell+15\right) T^2
+3 \delta ^3 \left(76 \ell^2+8 \ell+5\right) T
+6 \delta ^4 (\ell+2)\Big)\frac{C^{(1)}_\ell\left(\frac{2 T}{\delta }+1\right)}{24 \delta ^2 (\ell+1) T (\delta +T)^3}\nonumber\\
&-\frac{\left(6 \delta ^3+\left(4 \ell^2+8 \ell+42\right) T^3+\delta  \left(4 \ell^2+8 \ell+105\right) T^2+84 \delta ^2 T\right)C^{(1)}_{\ell+1}\left(\frac{2 T}{\delta }+1\right) }{24 \delta  T (\delta +T)^3}\bigg]\nonumber\\
&+\sum\limits_{\ell=0}^{\delta-1} \bigg[ \langle f_0 \tau_2 \rangle_{\delta ,\ell} \left(\frac{(-\delta  \ell-2 (\ell+1) T) C^{(1)}_\ell\left(\frac{2 T}{\delta }+1\right)}{\delta ^{3/2} (\ell+1) (\delta
   +T)}+\frac{C^{(1)}_{\ell+1}\left(\frac{2 T}{\delta }+1\right)}{\sqrt{\delta } (\delta +T)}\right)
- \frac{\langle f_2 \rangle_{\delta ,\ell} C^{(1)}_\ell\left(\frac{2 T}{\delta }+1\right)}{\delta +\delta \ell}\bigg]\,.
\nonumber
\end{align}

\section{Localisation constraints}
\label{loc}

In this appendix we will give the details of how to apply the localisation constraints of \cite{Behan:2023fqq,Chester:2022sqb} to the correlator for the $G_F=SO(8)$ theory. Of the three localisation constraints considered in \cite{Behan:2023fqq,Chester:2022sqb}, only one applies to the flavour structures we consider in \eqref{trace_basis}, and this constraint takes the form
\es{intNew}{
-\partial_{\mu_1}^4 F\big|_{\mu=0}+3\partial_{\mu_1}^2\partial_{\mu_2}^2 F\big|_{\mu=0}&=32 N^2I[M(s,t)+M(t,u)+M(u,s)]\,,\\
}
where the integral is defined as\footnote{We use shifted Mellin variables compared to \cite{Behan:2023fqq}: $(s,t,u)_\text{here} = (s-2,t-2,{u}-2)_\text{there}$.}
\es{Imel}{
  I[M]\equiv -\int \frac{ds dt}{(4\pi i)^2}& \Bigg[M(s,t) \Gamma[1-s/2]\Gamma[1+s/2]\Gamma[1-t/2]\Gamma[1+t/2]\Gamma[1-{u}/2]\Gamma[1+{u}/2] \\
 & \times \Big(\frac{H_{\frac s2}+H_{-\frac s2}}{tu}+\frac{H_{\frac t2}+H_{-\frac t2}}{su}+\frac{H_{\frac u2}+H_{-\frac u2}}{st}\Big)\Bigg]\,,
}
where $H_n$ is a harmonic number. The LHS of \eqref{intNew} is written in terms of derivatives of the free energy deformed by two of the four masses $\mu_i$ corresponding to the four Cartans of $SO(8)$. This quantity was computed in a large $N$ and large $\lambda$ expansion in \cite{Behan:2023fqq}, and takes the form
\es{intNew2}{
-\partial_{\mu_1}^4 F\big|_{\mu=0}+3\partial_{\mu_1}^2\partial_{\mu_2}^2 F\big|_{\mu=0}&=\frac{192\zeta(2)}{\lambda}N+O(N^0)\,.\\
}
For the Mellin amplitude to the order shown in \eqref{melAns}, we consider the integrals\footnote{Note that the meromorphic term cancels out from the combination in \eqref{intNew}.}
\es{intsApp}{
I[1]&=\frac{1}{24}\,,\qquad I[s]=I[u]=I[t]=0\,,\qquad I[t^2]=I[s^2]=I[u^2]=-\frac{1}{30}\,,\\
I[st]&=I[su]=I[st]= \frac{1}{60}\,.
}
Applying \eqref{intNew} to \eqref{melAns} using these integrals and the localisation result \eqref{intNew2}, we get \eqref{localisation}.

\section{Low energy expansion for integrals with MPLs}
\label{app:LEE}

In this appendix we explain how to compute the low energy expansion around the point $S = T = 0$ for integrals of the form
\beq
I_w (S,T) = \int\limits_0^1 dz \, z^{-S-1} (1-z)^{-T-1} L_w (z)\,.
\eeq
This computation is a simpler version of the analogous two-dimensional case done for closed string amplitudes in \cite{Vanhove:2018elu}.
To deal with the singularities of the integrand at $z=0, 1$ we start by splitting the integral into two contributions
\beq
I_w (S,T) = I^{(1)}_w (S,T) + \lim\limits_{\e \to 0} I^{(2)}_w (S,T)\,,
\eeq
where
\bea
I^{(1)}_w (S,T) &= \int\limits_0^1 dz \, \frac{\left(z^{-S} -1\right) \left((1-z)^{-T}-1\right)}{z(1-z)} L_w (z)\,,\\
I^{(2)}_w (S,T) &= \int\limits_\e^{1-\e} dz \, \frac{z^{-S} +(1-z)^{-T}-1}{z(1-z)} L_w (z)\,.
\eea{I12}
The first contribution is absolutely convergent at $S=T=0$ so that we can first Taylor expand around this point and then integrate term by term. Using the shuffle relations
\beq
 L_w (z)  L_{w'} (z) = \sum\limits_{W \in w \shuffle w'}  L_W (z)\,,
\label{shuffle}
\eeq
and
\beq
\frac{d}{dz} \left( L_{0W}(z) - L_{1W}(z) \right) = \frac{L_W(z)}{z(1-z)}\,,
\eeq
we find
\bea
I^{(1)}_w (S,T) &= \sum\limits_{p,q=1}^{\infty} (-S)^p (-T)^q \int\limits_0^1 dz \,
\frac{L_{0^p}(z) L_{1^q}(z) L_{w}(z)}{z(1-z)}\\
&=\sum\limits_{p,q=1}^{\infty} (-S)^p (-T)^q \sum\limits_{W \in 0^p \shuffle 1^q \shuffle w} \left(L_{0W}(1) - L_{1W}(1) \right)\,.
\eea{I1}
The second contribution is also absolutely convergent at $S=T=0$ as long as $\e>0$, so we compute
\beq
I^{(2)}_w (S,T) = \sum\limits_{\substack{p,q=0\\p\cdot q =0}}^{\infty} (-S)^p (-T)^q 
\sum\limits_{W \in 0^p \shuffle 1^q \shuffle w} 
\int\limits_\e^{1-\e} dz \, \frac{L_W(z)}{z(1-z)}\,,
\eeq
where the integral gives
\beq
\int\limits_\e^{1-\e} dz \, \frac{L_W(z)}{z(1-z)}
= L_{0W}(1-\e) - L_{1W}(1-\e) - \left(L_{0W}(\e) - L_{1W}(\e) \right)\,.
\eeq
The contribution from $z=0$ always vanishes, except when $w=0^n$. In this case it contributes
\beq
\lim\limits_{\epsilon \to 0} \sum_{p=0}^\infty (-S)^p \binom{p+n}{n} \frac{-\log^{p+n+1}(\e)}{(p+n+1)!}
= -\frac{1}{S^{n+1}}\,.
\eeq
The contribution from $z=1$ is more subtle, as MPLs $L_{w}(z)$ can have a logarithmic singularity near $z=1$ when the last letter in $w$ is 1. These can be isolated in terms of 
$L_{1^n}(1-\e) = \frac{1}{n!} \log^n(\e)$ using the shuffle relations \eqref{shuffle}.
Assume for instance $w=w'01$. In this case we can use
\beq
L_{w'0}(z) L_1(z) = L_w(z) + \sum\limits_{W \in w' \shuffle 1} L_{W0}(z)\,,
\eeq
to isolate the singularity
\beq
L_w(1-\e) = L_{w'0}(1) \log(\e) - \sum\limits_{W \in w' \shuffle 1} L_{W0}(1) + O(\e)\,.
\eeq
This idea can be used recursively to determine the singular contributions near $z=1$ for any word $w$, which lead to poles in $T$.
In general the integral then takes the form
\beq
\lim\limits_{\epsilon \to 0} I^{(2)}_w (S,T) = \text{poles } + 
\sum\limits_{\substack{p,q=0\\p\cdot q =0}}^{\infty} (-S)^p (-T)^q 
\sum\limits_{W \in 0^p \shuffle 1^q \shuffle w} 
\left( L_{0W}(1) - L_{1W}(1) \right)\,,
\eeq
and the sum of both contributions gives
\beq
I_w (S,T) = \text{poles } + 
\sum\limits_{p,q=0}^{\infty} (-S)^p (-T)^q 
\sum\limits_{W \in 0^p \shuffle 1^q \shuffle w} 
\left( L_{0W}(1) - L_{1W}(1) \right)\,.
\eeq

\section{Classical string solutions} \label{apd:classical}

In this appendix we give more details about the classical string solutions of Section \ref{class}. We first review the basics of classical string solutions in $AdS_5 \times S^5$, then give a summary of the open string solutions we found and finally give a detailed description of the folded closed string solutions and the corresponding open string solutions.

\subsection{Setup}

The range of $\sigma$ for the closed strings is within $[0,2\pi)$. For open strings this range is halved, resulting in $\sigma \in [0,\pi)$.

\paragraph{Action.}

The bosonic part of the $AdS_5 \times S^5$ string action in conformal gauge reads
\beq \label{eqn-ActionString}
I_{B}= \frac{\sqrt{\lambda}}{4\pi} \int d\tau \int {d\sigma} 
 \ (  L_{AdS} + L_S )\ ,  
\eeq
where
\beq \label{eqn-ActionString-02}
L_{AdS} =- \partial_a Y_P \partial^a Y^P - 
\tilde \Lambda (Y_P Y^P+1)\, , \  \ \ \ \ \ \
L_S =- \partial_a X_M\partial^a X_M+ \Lambda 
(X_M X_M-1)\,  .    
\eeq
Here $X_M$, $M=1,\ldots , 6$ and $Y_P$, 
 $P=0,\ldots , 5$ are  
the embedding coordinates of $\mathbb{R}^6$ 
with the Euclidean metric $\delta_{MN}$ and of $\mathbb{R}^{2,4}$ 
 with  $\eta_{PQ}=(-1,+1,+1,+1,+1,-1)$ 
 in $L_{AdS}$,  respectively  ($Y_P = \eta_{PQ} Y^Q$). 
 $\Lambda$  and $\tilde \Lambda$ are the Lagrange multipliers
 imposing the two hypersurface conditions. 

 \paragraph{Global AdS Coordinates.}

 It is conventional to use global coordinates, defined as
 \beq \label{eqn-GlobalCoord}
 \begin{aligned}
 Y_1 + i Y_2 & = \sinh \rho \sin \theta e^{i \phi_1}, \quad & X_1 + i X_2 &= \sin \gamma \cos \psi e^{i\varphi_1}, \\ 
 Y_3 + i Y_4 & = \sinh \rho \cos \theta e^{i \phi_2}, \quad & X_3 + i X_4 & = \sin \gamma \sin \psi e^{i\varphi_2}, \\
 Y_5 + i Y_0 & = \cosh \rho e^{i \phi_3}, \quad & X_5 + i X_6 & = \cos \gamma e^{i\varphi_3}\, . \\
 \end{aligned}
 \eeq
 The metric reads
 \beq \label{eqn-GlobalCoord-Metric}
\begin{aligned}
    (\dd s^2)_{AdS_5} & = \dd \rho^2 - \cosh \rho^2 \dd t^2 + \sinh^2 \rho (\dd \theta^2 + \cos^2 \theta \dd \phi_1^2 + \sin^2 \theta \dd \phi_2^2 )\, ,\\
    (\dd s^2)_{S^5} & = \dd \gamma^2 + \cos \gamma^2 \dd \varphi_3^2 + \sin^2 \gamma (\dd \psi^2 + \cos^2 \psi \dd \varphi_1^2 + \sin^2 \psi \dd \varphi_2^2 )\, .                         \\
\end{aligned}
\eeq
 
 \paragraph{Equations of Motion.}
 The classical equations of motion read
 \beq
 \begin{aligned} \label{eqn-apd-EOM}
 \partial^a \partial_a Y_P -  \tilde \Lambda  Y_ P &=0 \ , &\quad
 \tilde \Lambda &=    \partial^a  Y_P \partial_a Y^P  \ ,&\quad
 Y_P Y^P&=- 1\ ,\\
 \partial^a \partial_a X_M   +  \Lambda  X_M  &=0 \ ,&\quad
 \Lambda  &= \partial^a X_M  \partial_a  X_M  \ , &\quad
X_M X_M &=1 \ . 
 \end{aligned}
 \eeq  
In addition, the coordinates satisfy Virasoro constraints. In conformal gauge they read 
 \beq \label{eqn-apd-Virso}
 \dot{Y}_P\dot{Y}^P+  Y'_P  Y'^P
 +   \dot{X}_M\dot{X}_M+ X_M ' X_M ' =0 \ , \ \ \ \ \ \ \
   \dot{Y}_P Y'^P+ \dot X_M  X_M '  =0 \ . 
\eeq

\paragraph{Conserved Charges.}

The corresponding $SO(2,4)$ and $SO(6)$  conserved charges are
\footnote{These charges are canonically normalised. The reason is the canonical momenta of $X_M$ in terms of the normalisation of the action \eqref{eqn-ActionString} is given by $P_M = \frac{\partial L}{\partial \dot{X}_M} = \sqrt{\lambda}/2\pi$, so $J_{MN} = \int d\sigma (X_M P_N - X_N P_M)$. The factor $1/2\pi$ is unrelated to the range of $\sigma$.}
 \beq 
 {\bf S}_{PQ}= \sqrt{\lambda } \int {d\sigma \over 2\pi }
\ (Y_P \dot Y_Q  -  Y_Q\dot Y_P)\, ,\quad
J_{MN}= \sqrt{\lambda } \int {d\sigma \over 2\pi }
\ (X_M\dot  X_N  -  X_N\dot X_M)
\eeq
The conventional choice for the 3+3 Cartan generators of
$SO(2,4) \times SO(6)$ is
\beq \label{apd-ConservedCharges}
\begin{aligned}
{\bf S}_0 \equiv  {\bf S}_{50} \equiv E = \sqrt{\lambda} \mathcal{E} \ , \ \ \ \  {\bf S}_1 \equiv
{\bf S}_{12}= 
\sqrt{\lambda} \mathcal{S}_1 \ , \ \
\  {\bf S}_2\equiv {\bf S}_{34} =  \sqrt{\lambda} \mathcal{S}_2\ , \\
J_1 \equiv  J_{12} = \sqrt{\lambda} \mathcal{J}_1, \quad
J_2 \equiv  J_{34}  =\sqrt{\lambda} \mathcal{J}_2, \ \ \ \ \ 
J_3 \equiv  J_{56}= \sqrt{\lambda}\mathcal{J}_3 \ .
\end{aligned}
\eeq

\paragraph{Boundary Conditions.}

For closed string solutions we need to impose periodic boundary conditions for all the fields.
For open string solutions, the bosonic fields $Y_P$ in $AdS_5$ all satisfy the Neumann boundary conditions. Out of three bosonic fields $X_M$ in $S^5$, two satisfy Neumann boundary conditions and one satisfies Dirichlet boundary conditions.

\subsection{Classical string solutions}

The classical solutions are labelled by their quantum numbers $({\bf S}_1, {\bf S}_2|J_1,J_2,J_3)$ defined in \eqref{apd-ConservedCharges}.
There is a very simple relation to get the \textit{classical} open string energy from the closed one, as long as the restriction of the closed string solution to half of its range leads to a valid open string solution that satisfies the correct boundary conditions \cite{Stefanski:2003qr}. It is given by
\beq \label{eqn-ClassicalRel}
E_\text{cl}^{\text{open}}({\bf S}_1^{\text{open}},{\bf S}_2^{\text{open}}|J_1^{\text{open}},J_2^{\text{open}},J_3^{\text{open}}) = \frac{1}{2} E_\text{cl}^{\text{closed}}(2{\bf S}_1^{\text{open}},2{\bf S}_2^{\text{open}}|2J_1^{\text{open}},2J_2^{\text{open}},2J_3^{\text{open}}) \, . 
\eeq

\subsection{Summary of open string solutions}

We manage to construct the following types of classical open-string solutions based on the known closed string solutions in the literature \cite{Gubser:2002tv,Minahan:2002rc,Frolov:2002av,Frolov:2003qc,Frolov:2003tu,Arutyunov:2003za,Beisert:2003ea,Tirziu:2008fk,Beccaria:2008dq,Roiban:2009aa,Roiban:2011fe,Frolov:2003xy,Tseytlin:2004xa,Tseytlin:2010jv}
\begin{itemize}
    \item  Circular $S^5$: the state has quantum numbers $(0,0|J_1,J_3,J_1)$ and satisfies $(N,N,D)$ boundary conditions on $S^5$ with classical energy 
    \beq
    E = \sqrt{2 m \sqrt{\lambda} J_1} \left( 1+ \frac{J_3^2}{4mJ_1 \sqrt{\lambda}} + \mathcal{O}\left(1/\lambda\right) \right)\, .
    \eeq
    \item Circular $AdS_5$: the state has quantum numbers $({\bf S},{\bf S}|J,0,0)$ and satisfies $(N,N,D)$ boundary conditions on $S^5$. However, the solution satisfies $(N,D)$ boundary conditions in the $AdS$ direction and thus does not correspond to a valid open string solution.
    \item Circular Mixed: the state has quantum numbers $({\bf S},0|{\bf S},0, J)$ and satisfies $(N,N,D)$ boundary conditions on $S^5$ with classical energy. This solution is a complex solution so restricting it to $\sigma \in [0,\pi]$ will not satisfy the needed boundary conditions.
    \item Folded Mixed: the state has quantum numbers $({\bf S},0|J,0,0)$ and satisfies $(N,N,D)$ boundary conditions on $S^5$ with classical energy 
    \beq \label{eqn-Dispersion-Fold-SJ-Open}
    E =  \sqrt{m \sqrt{\lambda} {\bf S}} \left(1 + \frac{1}{\sqrt{\lambda}} \left(\frac{3{\bf S}}{4m} + \frac{J^2}{2 {\bf S} m} \right)+ \mathcal{O}(1/\lambda) \right)\, .
    \eeq
     Details can be found in \eqref{eqn-Dispersion-Fold-SJ}. To get an $SU(2)_L$ singlet state, we must take $J =0$.
    \item Folded $S^5$: the state has quantum numbers $(0,0|J_1,J_2,0)$ and satisfies $(N,N,D)$ boundary conditions on $S^5$ with classical energy 
    \beq \label{eqn-Dispersion-Fold-JJ-Open}
    E =  \sqrt{m \sqrt{\lambda} J_2} \left(1 + \frac{1}{ \sqrt{\lambda}} \left(\frac{J_2}{4 m} + \frac{J_1^2}{2 m J_2}\right)+ \frac{1}{ \lambda} \left(-\frac{J_1^4}{8 J_2^2 m^2}+\frac{7 J_1^2}{8 m^2}+\frac{3 J_2^2}{32 m^2}\right) + \mathcal{O}(\lambda^{-\frac{3}{2}})\right).
    \eeq
    To get an $SU(2)_L$ singlet state we have to take $J_1 = J_2 = J$.  Details can be found in \eqref{eqn-Dispersion-Fold-JJ}.
    \item Glued Folded: the state is obtained by gluing the two folded solutions presented above. The state has quantum numbers $({\bf S},0|J_1,J_2,0)$ and satisfies $(N,N,D)$ boundary conditions on $S^5$ with classical energy
\begin{align}
        E ={}&  \sqrt{m\sqrt{\lambda} ({\bf S}+J_1)} \bigg(1 + \frac{1}{\sqrt{\lambda} m} \left(\frac{3{\bf S}}{4} + \frac{J_1}{4} +\frac{J^2_2}{2({\bf S}+J_1)} \right) \\
        &+ \frac{1}{\lambda m^2} \left(\frac{3 J_1^2}{32}-\frac{J_2^4}{8 (J_1+{\bf S})^2}+\frac{J_1 J_2^2}{4 (J_1+{\bf S})}-\frac{13 J_1 {\bf S}}{16}+\frac{5 J_2^2}{8}-\frac{21 {\bf S}^2}{32}\right)  + \mathcal{O}(1/\lambda^\frac{3}{2}) \bigg)\, .\nonumber
    \end{align}
\end{itemize}

\subsection{Folded closed string solutions}

We first present the classical closed folded string solutions in \cite{Frolov:2003xy}.

\subsubsection{Mixed folded solution: $({\bf S}, 0| J, 0, 0)$ type}

\begin{figure}[!ht]
    \centering
    \includegraphics[width= 0.3 \textwidth]{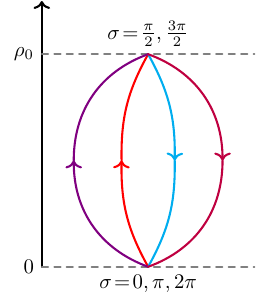}
    \caption{Illustration of the closed mixed folded string solution. The solution contains four identical segments, which are split in the plot for clarity. The four segments correspond to \(\sigma\) in the intervals \([0, \pi/2]\), \([\pi/2, \pi]\), \([\pi, 3\pi/2]\), and \([3\pi/2, 2\pi]\), respectively, and are coloured in \textcolor{red}{red}, \textcolor{cyan}{cyan}, \textcolor{violet}{violet}, and \textcolor{purple}{purple} for visual distinction.}
    \label{fig:FoldedClosedAdS}
\end{figure}

For the mixed folded solution, the coordinates take the following form,
\beq \label{eqn-apd-folded-SJCoord}
\theta =0, \quad \rho = \rho(\sigma), \quad \phi_3 = k \tau, \quad \gamma=0, \quad \phi_2 = \omega \tau, \quad \varphi_3 =\nu \tau.
\eeq
The EOMs and the Virasoro constraints read
\beq
\begin{aligned}
\rho'' + \frac{1}{2} (\omega_{}^2-k^2) \sin 2\rho & = 0, \\
(\rho')^2 + \omega^2 \sinh^2 \rho - k^2 \cosh^2 \rho & = \nu^2 \, .
\end{aligned}
\eeq
The function $\rho$ consists of four identical parts, as depicted in figure \ref{fig:FoldedClosedAdS}. For $\sigma \in [0,\pi/2]$ $\rho(\sigma)$ increases from $\rho =0$ to its maximal value $\rho_0$ where $\rho' =0$. In the next interval  $\sigma \in [\pi/2,\pi]$, $\rho(\sigma)$ reverses direction, decreasing back to zero along the same path. This pattern repeats for $\sigma \in [\pi,3\pi/2]$ and $\sigma \in [3\pi/2, 2\pi]$.
Since the maximal value $\rho_0$ of $\rho$ is obtained when $\rho' =0$, we can express $\rho_0$ as 
\beq
\omega^2 \sinh^2 \rho_0 - k^2 \cosh^2 \rho_0 = \nu^2\, .
\eeq
Substituting $\nu^2$ to the 2nd equation above, we find,
\beq
\frac{\dd \rho}{\dd \sigma} = \sqrt{\omega^2- k^2} \sqrt{\sinh^2 \rho_0 - \sinh^2 \rho}\, ,
\eeq
for $\sigma \in [0,\pi/2]$ and $\rho \in [0,\rho_0]$. This differential equation can be solved in terms of the Jacobi Elliptic functions but we will not present it here.

For consistency, the total change of $\sigma$ is given by $2\pi$, which leads to the constraint
\beq \label{eqn-Dispersion-Fold-m}
2\pi = \int_0^{2\pi} \dd \sigma = \frac{4m}{\sqrt{\omega^2 -k^2}} \int_0^{\rho_0} \dd \rho \frac{1}{\sqrt{\sinh^2 \rho_0 - \sinh^2 \rho}} \, .
\eeq
Here we have introduced an integer parameter $m$ which counts the total number of times that a string folds. The solution in figure \ref{fig:FoldedClosedAdS} corresponds to $m=1$.

\paragraph{Energy.}

The physical state has energy
\beq \label{apd-folded-SJ-E}
E = \frac{\sqrt{\lambda}}{2\pi} k \int \dd \sigma \cosh^2 \rho = \frac{\sqrt{\lambda}}{2\pi} k \frac{4m}{\sqrt{\omega^2 -k^2}} \int_0^{\rho_0} \dd \rho \frac{\cosh^2 \rho}{\sqrt{\sinh^2 \rho_0 - \sinh^2 \rho}}\,,
\eeq
and carries spin and angular momentum
\beq \label{apd-folded-SJ-SJ}
\begin{aligned}
{\bf S} & = \frac{\sqrt{\lambda}}{2\pi} \omega \int \dd \sigma \sinh^2 \rho = \frac{\sqrt{\lambda}}{2\pi} \omega \frac{4m}{\sqrt{\omega^2 -k^2}} \int_0^{\rho_0} \dd \rho \frac{\sinh^2 \rho}{\sqrt{\sinh^2 \rho_0 - \sinh^2 \rho}}\,, \\
J & = \frac{\sqrt{\lambda}}{2\pi} \nu \int \dd \sigma  = \sqrt{\lambda} \nu  \, .
\end{aligned}
\eeq
Solving the constraints for the energy we find
\beq \label{eqn-Dispersion-Fold-SJ}
E =  \sqrt{2 m \sqrt{\lambda} {\bf S}} \left(1 + \frac{1}{\sqrt{\lambda}} \left(\frac{3{\bf S}}{8 m } + \frac{J^2}{4 m {\bf S}} \right)+ \mathcal{O}(1/\lambda) \right)\, .
\eeq
The energy can also be computed in other limits, in particular, in the $u = \mathcal{J}^2/\mathcal{S}$ fixed limit. It reads,
\beq \label{eqn-Dispersion-Fold-SJ-limit}
\mathcal{E} = \sqrt{u+2} \sqrt{\mathcal{S}} \left(1 + \frac{2u+3}{4(u+2)} \mathcal{S} -\frac{2 u[2 u(u+5)+17]+21}{32(u+2)^2} \mathcal{S}^{2} \right)+\ldots \,,
\eeq
where we recall that $\mathcal{E} = E/\sqrt{\lambda}$, $\mathcal{S} = {\bf S}/\sqrt{\lambda}$ and $\mathcal{J} = J/\sqrt{\lambda}$.

\subsubsection{Folded $S^5$ solution: $(J_1, J_2, 0)$ type}

For the folded $S^5$ solution, the coordinates take the following form,
\beq \label{eqn-apd-folded-JJCoord}
\rho =0, \quad \phi_3 = k \tau, \quad \gamma = \frac{\pi}{2}, \quad \psi = \psi(\sigma), \quad \varphi_1 =\omega_1 \tau, \quad \varphi_2 = \omega_2 \tau.
\eeq
The EOM and the Virasoro constraints read,
\beq
\begin{aligned}
\psi'' + \frac{1}{2} \omega_{21}^2 \sin 2\psi & = 0, \\
(\psi')^2 + \omega_1^2 \cos \psi^2 + \omega_2^2 \sin^2 \psi & = k^2 \, ,
\end{aligned}
\eeq
where $\omega_{21}^2 \equiv \omega_2^2 - \omega_1^2 \geq 0$. 

\begin{figure}[!ht]
    \centering
    \includegraphics[width= 0.4 \textwidth]{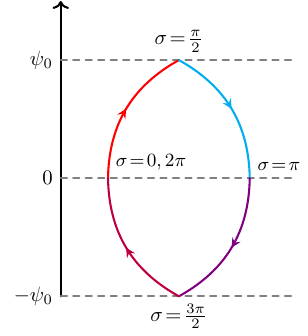}
    \caption{Illustration of the closed folded $S^5$ string solution. The solution contains four identical segments, which are split in the plot for clarity. The four segments correspond to \(\sigma\) in the intervals \([0, \pi/2]\), \([\pi/2, \pi]\), \([\pi, 3\pi/2]\), and \([3\pi/2, 2\pi]\), respectively, and are coloured in \textcolor{red}{red}, \textcolor{cyan}{cyan}, \textcolor{violet}{violet}, and \textcolor{purple}{purple} for visual distinction.}
    \label{fig:FoldedClosedS}
\end{figure}

The function $\psi$ consists of four identical parts, as depicted in figure \ref{fig:FoldedClosedS}. For $\sigma \in [0,\pi/2]$ $\psi(\sigma)$ increases from $\psi =0$ to its maximal value $\psi_0$ where $\psi' =0$. In the next interval $\sigma \in [\pi/2,\pi]$, $\psi(\sigma)$ reverses direction, decreasing back to zero along the same path. With further increase in $\sigma$ within $[\pi, 3\pi/2]$, $\psi(\sigma)$ declines from zero to its minimum at $-\psi_0$, then increases back to zero as $\sigma$ extends from $3\pi/2$ to $2\pi$.
Since the maximal value $\psi_0$ of $\psi$ is obtained when $\psi' =0$, we can express $\psi_0$ as 
\beq
\omega_1^2 \cos \psi^2_0 + \omega_2^2 \sin^2 \psi_0  = k^2\, .
\eeq
Substituting $k^2$ to the 2nd equation above, we find, 
\beq
\frac{\dd \psi}{\dd \sigma} = \omega_{21} \sqrt{\sin^2 \psi_0 - \sin^2 \psi}\, .
\eeq

For consistency, the total change of $\sigma$ is given by $2\pi$, thus leads to the constraint
\beq  \label{eqn-Dispersion-Fold2-m}
2\pi = \int_0^{2\pi} \dd \sigma = 4 m \int_0^{\psi_0} \dd \psi \frac{1}{\omega_{21} \sqrt{\sin^2 \psi_0 - \sin^2 \psi}}\, .
\eeq
Here we have introduced an integer parameter $m$ which counts the total number of times that a string folds. The solution in figure \ref{fig:FoldedClosedS} corresponds to $m=1$.

\paragraph{Energy.}

The physical state has energy
\beq \label{eqn-Dispersion-Fold2-E}
E = \sqrt{\lambda} k = \sqrt{\lambda} \sqrt{\omega_1^2 \cos^2 \psi_0 + \omega_2^2 \sin^2 \psi_0 }\,,
\eeq
and carries angular momenta
\beq
\begin{aligned} \label{eqn-Dispersion-Fold2-JJ}
J_1 & = \sqrt{\lambda} \omega_1 \int \frac{\dd \sigma}{2\pi} \cos^2 \psi = \sqrt{\lambda} \omega_1 \frac{4 m}{2\pi} \int_0^{\psi_0} \dd \psi \frac{\cos^2 \psi}{\omega_{21} \sqrt{\sin^2 \psi_0 - \sin^2 \psi}}\,, \\
J_2 & = \sqrt{\lambda} \omega_2 \int \frac{\dd \sigma}{2\pi} \cos^2 \psi = \sqrt{\lambda} \omega_2 \frac{4 m}{2\pi} \int_0^{\psi_0} \dd \psi \frac{\sin^2 \psi}{\omega_{21} \sqrt{\sin^2 \psi_0 - \sin^2 \psi}} \,.
\end{aligned}
\eeq
Solving for the energy, we find 
\beq \label{eqn-Dispersion-Fold-JJ}
E =  \sqrt{2 m \sqrt{\lambda} J_2} \left(1 + \frac{1}{ \sqrt{\lambda}} \left(\frac{J_2}{8 m} + \frac{J_1^2}{4 m J_2}\right)+ \frac{1}{ \lambda} \left(-\frac{J_1^4}{32 J_2^2 m^2}+\frac{7 J_1^2}{32 m^2}+\frac{3 J_2^2}{128 m^2}\right) + \mathcal{O}(\lambda^{-\frac{3}{2}})\right)\, .
\eeq

\subsubsection{Glued solution}

One can construct a more generic solution by taking the $AdS_5$ part of \eqref{eqn-apd-folded-SJCoord} and the $S^5$ part of \eqref{eqn-apd-folded-JJCoord}. The coordinate take the following form
\beq
\theta =0, \quad \rho = \rho(\sigma), \quad \phi_3 = k \tau, \quad \gamma = \frac{\pi}{2}, \quad \psi = \psi(\sigma), \quad \varphi_1 =\omega_1 \tau, \quad \varphi_2 = \omega_2 \tau.
\eeq
and the functions $\rho$ and $\psi$ are taken to be the same as introduced in the previous subsections.

Notice that the $AdS_5$ and $S^5$ EOMs (see \eqref{eqn-apd-EOM}) are decoupled, so the ansatz above solves the EOMs automatically. The only non-trivial constraint comes form the Virasoro constraints \eqref{eqn-apd-Virso}. One immediately notices that the Virasoro constraints are solved once we identify
\beq
\nu^2|_{(S,J)} = k^2|_{(J,J')}\, .
\eeq
As a result, the final expression for the energy is obtained by using the $(S,J)$-type energy expression and identifying
\beq
J^2|_{(S,J)} = E^2|_{(J,J')}\, .
\eeq
In practice, we only need to identify the RHS of \eqref{eqn-Dispersion-Fold-JJ} as $\mathcal{J}$ in \eqref{eqn-Dispersion-Fold-SJ-limit} and re-expand, we immediately find
\bea
    E ={}&  \sqrt{m\sqrt{\lambda} ({\bf S}+J_1)} \bigg(1 + \frac{1}{m\sqrt{\lambda}} \left(\frac{3{\bf S}}{8} + \frac{J_1}{8} +\frac{J^2_2}{4({\bf S}+J_1)} \right)\\
    &+\frac{1}{\lambda m^2} \left(\frac{3 J_1^2}{8}-\frac{J_2^4}{2 (J_1+{\bf S})^2}+\frac{J_1 J_2^2}{J_1+{\bf S}}-\frac{13 J_1 {\bf S}}{4}+\frac{5 J_2^2}{2}-\frac{21 {\bf S}^2}{8}\right)  + \mathcal{O}(1/\lambda^\frac{3}{2}) \bigg)\,.
\eea{eqn-Dispersion-Fold-Glued}

\subsection{Folded open string solutions}

Finding open string solutions from the known closed ones was initiated in \cite{Stefanski:2003qr,Chen:2004yf,Okamura:2005cj}. We will see that valid open string solutions can be obtained by properly choosing the segments of the closed string solutions, and thus the energy is simply obtained by using the identification \eqref{eqn-ClassicalRel}.

It is important to remember, as discussed in section \ref{sec:classicalOpenCond}, that the boundary conditions for open strings were fixed: all fields in $AdS$ along with four fields in $S^5$ satisfy the Neumann boundary conditions, whereas two fields in $S^5$ satisfy with the Dirichlet boundary conditions.

\paragraph{Mixed folded string.}

\begin{figure}[h]
    \centering
    \includegraphics[width= 0.3 \textwidth]{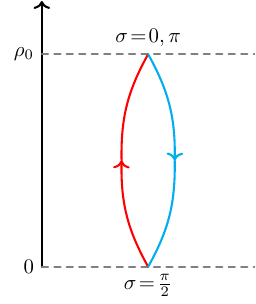}
    \caption{Open folded string solution from closed folded string solution: we take half of the solution and relabel the coordinate by $\sigma \rightarrow \sigma - \pi/2$.}
    \label{fig:FoldedOpenAdS}
\end{figure}

For the mixed folded $(S,J)$-type open string, we need the solution in all the AdS directions to satisfy the Neumann boundary conditions. To ensure this, we select two segments of the folded closed string that align with the Neumann boundary conditions at $\sigma = 0, \pi$, as shown in figure \ref{fig:FoldedOpenAdS}, where we recall that at the maximum $\rho|_{\sigma =0, \pi} = \rho_0 \neq 0$ and $\rho'|_{\sigma =0, \pi}=0$.
Comparing this with the global coordinates in \eqref{eqn-GlobalCoord}, we immediately find that all AdS coordinates $Y_P$ satisfy the Neumann boundary conditions, $Y_P'|_{\sigma=0,\pi}=0$. 

On the $S^5$ side, the $X_5, X_6$ fields satisfy Neumann boundary conditions, 
$X_5'|_{\sigma=0,\pi}= X_6'|_{\sigma=0,\pi}= 0$. Since $\gamma=0$, we have $X_1 + \ii X_2= X_3 + \ii X_4 \equiv 0$. Consequently, these directions can be considered as satisfying either Dirichlet or Neumann conditions. 

Now that we verified the boundary conditions we can compute the new energy. However, notice that the only change in the equations \eqref{apd-folded-SJ-E}, \eqref{apd-folded-SJ-SJ} and \eqref{eqn-Dispersion-Fold-m} for the open case is $4m \rightarrow 2m$ as the only change is the total number of identical segments. This observation immediately leads to the identification \eqref{eqn-ClassicalRel} and thus \eqref{eqn-Dispersion-Fold-SJ-Open}.

\paragraph{$S^5$ folded string.}

For the $S^5$ $(J_1,J_2)$-type folded solution, the Neumann boundary conditions in all AdS directions are manifest because $\rho =0$.

\begin{figure}[!ht]
    \centering
    \includegraphics[width= 0.3 \textwidth]{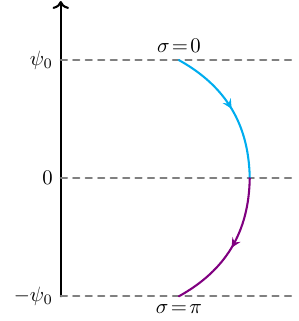}
    \caption{Open folded string solution from closed folded string solution: we take half of the solution and relabel the coordinate by $\sigma \rightarrow \sigma - \pi/2$.}
    \label{fig:FoldedOpenS}
\end{figure}

In the $S^5$ directions, the solution for $\psi$ can similarly be constrained to ensure that at $\sigma = 0, \pi$, we have $\psi \neq 0$ and $\psi' = 0$. This approach is illustrated in figure \ref{fig:FoldedOpenS}, where we take half of the closed string solution and modify the coordinate system by relabelling $\sigma \rightarrow \sigma - \pi/2$. This configuration leads to the application of mixed $(NND)$ boundary conditions in the $(12,34,56)$ directions, $X_1'|_{\sigma=0,\pi}= X_2'|_{\sigma=0,\pi}= X_3'|_{\sigma=0,\pi}= X_4'|_{\sigma=0,\pi}= X_5|_{\sigma=0,\pi} = X_6|_{\sigma=0,\pi} = 0$. Given that $X_5, X_6 \equiv 0$, these two directions can also be considered to satisfy the Neumann boundary conditions.

Now that we verified the boundary conditions we can compute the new energy. However, notice that the only change in the equations \eqref{eqn-Dispersion-Fold2-E}, \eqref{eqn-Dispersion-Fold2-JJ} and \eqref{eqn-Dispersion-Fold2-m} for the open case is $4m \rightarrow 2m$ as the only change is the total number of identical segments. This observation immediately leads to the identification \eqref{eqn-ClassicalRel} and thus \eqref{eqn-Dispersion-Fold-JJ-Open}.

\paragraph{Glued folded string.}

The construction of the glued folded string solution combines the previously discussed solutions for both the $AdS$ and $S^5$ sectors, which individually meet the requisite boundary conditions. Given this compatibility, it logically follows that the glued solution also satisfies to the appropriate boundary conditions. Consequently, the energy identified as \eqref{eqn-E-class-GluedFolded} is directly derived from \eqref{eqn-Dispersion-Fold-Glued} via the identification \eqref{eqn-ClassicalRel}.

\section{Holomorphic factorisation}
\label{facApp}
\begin{figure}[!ht]
\centering
{\includegraphics[height=1.5cm]{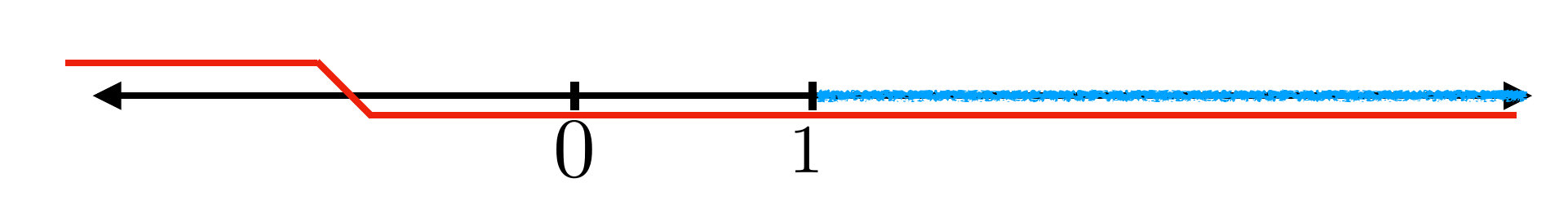}} 
\caption{For $z^+<0$ the integration contour, indicated in red, can be closed without crossing the branch cut, indicated in blue.}
\label{figf1}
\end{figure}

\begin{figure}[!ht]
\centering
{\includegraphics[height=1.5cm]{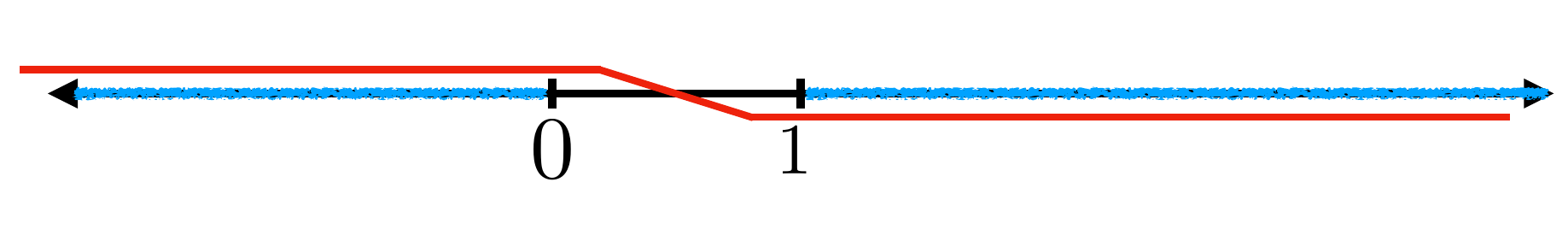}} 
\caption{For $z^+ \in [0,1]$ the integration contour, indicated in red, can be closed to wrap around either of the two branch cuts, indicated in blue.}
\label{figf2}
\end{figure}

In this appendix we show how to rewrite the result for the AdS Virasoro Shapiro amplitude \cite{Alday:2023mvu} as the product of one-dimensional integrals. We will review the usual KLT argument for holomorphic factorisation \cite{KAWAI19861}, and then explain how to generalise this, for the case of extra insertions of single-valued polylogarithms. This exercise will serve many purposes. On one hand it gives us some expectation as to which functions can appear in the open string case: not surprisingly, the usual, multi-valued multiple polylogarithms. On the other hand, it seems to be the best route to follow if one is interested in a closed form expression for the AdS Virasoro Shapiro in a curvature expansion. Consider as starting point the complex beta function - relevant for the computation of the usual VS amplitude, 
\begin{equation}
A_{\text{closed}}(S,T) =\int_{\mathbb{CP}^1} d^2 z |z|^{-2-2S}|1-z|^{-2-2T}\,,
\end{equation}
and write the integrand in terms of $x,y$ with $z=x+i y$ and $\bar z = x-i y$. We then make the following change of variables
\begin{equation}
y \to i e^{-2i \epsilon} y\,,
\end{equation}
where $\epsilon>0$ is a small positive number. It is convenient to introduce the following notation
\begin{equation}
z^{\pm} = x \pm y,~~~\delta = z^+ - z^-.
\end{equation}
Expanding to linear order in $\epsilon$ we then get
\begin{equation}
\label{almost}
A_{\text{closed}}(S,T) =\int_{-\infty}^\infty dz^+ dz^{-} (z^+- i \epsilon \delta)^{-S-1} (z^-+ i \epsilon \delta)^{-S-1}(z^+ -1- i \epsilon \delta)^{-T-1} (z^- -1 + i \epsilon \delta)^{-T-1}.
\end{equation}
This is almost factorised, but we need to be careful with branch cuts as we send $\epsilon \to 0$. For $x<0$ we choose
\begin{equation}
(x+ i \epsilon)^\alpha = e^{i \pi \alpha} (-x)^\alpha,~~~(x- i \epsilon)^\alpha = e^{-i \pi \alpha} (-x)^\alpha\,,
\end{equation}
while for $x>0$ we can of course take $\epsilon \to 0$ with no problem $(x \pm i \epsilon)^\alpha=x^\alpha$. 
Let's now fix $z^+$ to different values in (\ref{almost}), and then consider the integral over $z^-$:
\begin{equation}
A^{-}(z^+) =\int_{-\infty}^\infty dz^{-} (z^+- i \epsilon \delta)^{-S-1} (z^+ -1- i \epsilon \delta)^{-T-1} (z^-+ i \epsilon \delta)^{-S-1}(z^- -1 + i \epsilon \delta)^{-T-1}\,.
\end{equation}
Consider first $z^+<0$. The branch-points correspond to $z^- \sim 0$ and $z^- \sim1$. For both cases $\delta=z^+ - z^-<0$ and the integration contour lies below the real line at both branch points, see figure \ref{figf1}.
Note that the contour crosses the real line at $z^-=z^+<0$, but the integrand $ (z^+- i \epsilon \delta)^{-S-1} (z^-+ i \epsilon \delta)^{-S-1}$ is continuous as we cross the line. This means the integral vanishes, as we can close the contour without crossing any branch cuts. By the same reasoning, if $z^+>1$ then the integration contour for $z^-$ lies above the real line, and the integral again vanishes. Hence
\begin{equation}
A^{-}(z^+) =0 ~~~\text{for $z^+<0$},~~~A^{-}(z^+) =0 ~~~\text{for $z^+>1$}.
\end{equation}
This means that the only non-vanishing contribution arises from the region $z^+ \in [0,1]$ and we obtain 
\begin{equation}
A_{\text{closed}}(S,T) =\int_{0}^1 dz^+  (z^+)^{-S-1} (1-z^+)^{-T-1}  \int_{-\infty}^\infty dz^{-}  (z^-+ i \epsilon \delta)^{-S-1}(1-z^- - i \epsilon \delta)^{-T-1}\,,
\end{equation}
where we have written  $(z^+ -1- i \epsilon \delta)^{-T-1}(z^- -1 + i \epsilon \delta)^{-T-1}=(1-z^+ + i \epsilon \delta)^{-T-1}(1-z^- - i \epsilon \delta)^{-T-1}$. We now focus in the integral over $z^{-}$ and look at the behaviour around the branch points. For $z^- \sim 0$ we get $\delta = z^+ - z^->0$ and the contour is above the real line. For $z^- \sim 1$ we get $\delta = z^+ - z^-<0$ so that $1-z^- - i \epsilon \delta = 1-z^-_\epsilon$ where $z^-_\epsilon$ has a small negative imaginary part, and the contour is below the real line. In summary, we need to compute the following contour integral 
\begin{equation}
A_{1d}= \int_\gamma dz^{-}  (z^-)^{-S-1}(1-z^-)^{-T-1}\,,
\end{equation}
where $\gamma$ comes from $z=-\infty + i \epsilon$, goes below the real axis between $0$ and $1$ and goes to $z=\infty - i \epsilon$, see figure \ref{figf2}.
We can now close the contour, so that it goes around the real axis with $z>1$. The contribution to the integral arises then from the following discontinuity, for $x>1$

\begin{equation}
-\left. (1-z)^{-T-1} \right|_{z=x+i \epsilon} +\left. (1-z)^{-T-1} \right|_{z=x-i \epsilon} = 2i \sin(\pi T)(x-1)^{-T-1}\,,
\end{equation}
which leads to
\begin{equation}
A_{1d}=2i \sin(\pi T) \int_1^\infty dx  x^{-S-1}(x-1)^{-T-1}\,,
\end{equation}
and the KLT formula
\begin{equation}
A_{\text{closed}}(S,T) =2i \sin(\pi T)  \int_{0}^1 dz^+  (z^+)^{-S-1} (1-z^+)^{-T-1}   \int_1^\infty dx  x^{-S-1}(x-1)^{-T-1}\,.
\end{equation}

\subsection{Insertions}
We will now consider the extra insertion of single-valued multiple polylogarithms (SVMPLs)

\begin{equation}
A^{W}_{\text{closed}}(S,T) =\int_{\mathbb{CP}^1} d^2 z |z|^{-2S-2}|1-z|^{-2T-2}{\cal L}_W(z,\bar z).
\end{equation}
We now need to be careful with the manipulation of contours. In particular note that the change of coordinates performed above takes us away from the region where $\bar z = z^*$. For the insertions of interest in this paper, ${\cal L}_W(z,\bar z)$ can be written as a sum of products of multiple polylogarithms
\begin{equation} 
{\cal L}_W(z,\bar z) = \sum_{|w|+|w'|=|W|} c_{ww'} L_{w'}(z) L_w(\bar z) \,,
\end{equation}
where $w,w'$ are words in the $\{0,1\}$ alphabet. Each such term leads to the extra insertion $L_w(z^+) L_{w'}(z^-)$ after the change of variables considered above. Note that $L_w(z)$ contains branch cuts only along the real line, for $x<0$ and/or $x>1$. Because of this, the 'almost' factorised integral again vanishes for either $z^+<0$ or $z^+>1$ and one gets
\begin{equation}
A^{W}_{\text{closed}}(S,T)  =  \sum_{|w|+|w'|=|W|} c_{ww'}A^{w,w'}_{\text{closed}}(s,t)\,, 
\end{equation}
with 
\bea
A^{w,w'}_{\text{closed}}(S,T) ={}&\int_{0}^1 dz^+(z^+)^{-S-1} (1-z^+)^{-T-1}  L_w(z^+)  \\
&\times  \int_{-\infty}^\infty dz^{-}  (z^-+ i \epsilon \delta)^{-S-1}(1-z^- - i \epsilon \delta)^{-T-1} L_{w'}(z^-)\,.
\eea
We would like to stress that this identity is only true for the entire sum entering ${\cal L}_W(z,\bar z)$. Individual components would give non-vanishing contributions, for $z^+<0$ or $z^+>1$, which exactly cancel when considering the single-valued combination. Using this fact, the argument now runs exactly as for the case without insertions, except now we need to compute the discontinuity across the real axis for $x>1$ including the extra insertion
\begin{equation}
\text{Disc}_1[(1-z)^{-T-1} L_{w}(z)]\,.
\end{equation}
Multiple polylogarithms are generally not continuous as we cross the real axis for  $x>1$. For instance, for the simplest case of weight one
\begin{equation}
\text{Disc}_1[L_0(z)] = 0\,,~~~\text{Disc}_1[L_1(z)] =-2\pi i\,,
\end{equation}
where $\text{Disc}_1[L_w(x)]= L_w(x+i \epsilon)-L_w(x-i \epsilon)$ for $x>1$. For higher weights, let us introduce the following notation. Consider $L_w(x)$. In the region $x \in (0,1)$ this is well defined. Then

\bea
L_w(x \pm i \epsilon) &= L^+_w(x) \pm i \pi dL^+_w(x)\,,~~~\text{for $x>1$}\,, \\
L_w(x \pm i \epsilon) &= L^-_w(x) \pm i \pi dL^-_w(x)\,,~~~\text{for $x<0$}\,,
\eea{eqF1}
where $L^+_w(x)$ is well defined for $x>1$ and $L^-_w(x)$ is well defined for $x<0$. Then the relevant discontinuity to compute for the holomorphic factorisation is
\es{thing}{
&\text{Disc}_1[(1-z)^{-T-1} L_{w}(z)] \\
&=- e^{i \pi T} (z-1)^{-T-1} (L^+_w(x)+ i \pi dL^+_w(x))+e^{-i \pi T} (z-1)^{-T-1} (L^+_w(x)- i \pi dL^+_w(x)) \\
&=-2i (z-1)^{-T-1}  (\sin (\pi T) L^+_w(x) + \pi \cos (\pi T) dL^+_w(x))\,.
}
This leads to the final expression for the holomorphic factorisation with insertions:
\begin{align}
A^{w,w'}_{\text{closed}}(S,T)={}&2 i \int_{0}^1 dz^+(z^+)^{-S-1} (1-z^+)^{-T-1}  L_w(z^+)   \\
&  \times \int_{1}^\infty dz^{-} (z^{-} )^{-S-1} (z^{-} -1)^{-T-1}  (\sin (\pi T) L^+_{w'}(z^{-} ) + \pi \cos (\pi T) dL^+_{w'}(z^{-}))\,. \nonumber
\end{align}
\subsection{Relating integrals and symmetry}
The KLT relations are often written in a symmetric manner, with both integrals involved evaluated over the $(0,1)$ interval. To relate $1d$ integrals in different intervals, consider the following integrals, whose integration contours are parallel to the real line, with a small imaginary part, positive and negative respectively:
\begin{align}
e^{\mp i \pi (S+1)}(I_1^{w}  \pm i \pi dI_1^{w})  &= e^{\mp i \pi (S+1)} \int_{-\infty}^0 dz (-z)^{-S-1} (1-z)^{-T-1} \left( L^-_w(z) \pm i \pi dL^-_w(z) \right)\,, \nonumber\\
I^{w}_2 &= \int_{0}^1 dz (z)^{-S-1} (1-z)^{-T-1} L_w(x)\,, \label{more}\\
e^{\pm i \pi (T+1)} (I_3^{w}  \pm i \pi dI_3^{w})  &= e^{\pm i \pi (T+1)} \int_{1}^\infty dz (z)^{-S-1} (z-1)^{-T-1} \left(  L^+_w(z) \pm i \pi dL^+_w(z) \right)\,.\nonumber
\end{align}
Then we have 
\begin{equation}
e^{\mp i \pi (S+1)}(I_1^{w}  \pm i \pi dI_1^{w}) +I^{w}_2+e^{\pm i \pi (T+1)} (I_3^{w}  \pm i \pi dI_3^{w}) =0\,,
\end{equation}
as for both combinations we are able to close the contours without crossing any branch cuts. Consider the integrals with no insertion, so that $L_w(x)=1$ and no discontinuity. The above equations imply
\begin{equation}
 I_3 = \sin (\pi S) \csc (\pi  (S+T)) I_2\,.
\end{equation}
More generally note that $dI_1^{w},dI_3^{w}$ involve insertions of weight $|w|-1$, so that for instance
\begin{equation}
I^w_3 =  \sin (\pi S) \csc (\pi  (S+T)) I^w_2 + \text{lower weight insertions}\,.
\end{equation}
The precise lower weight insertions can be obtained by writing the discontinuities in terms of multiple polylogs of weight $|w|-1$, and so on. Let's now return to the specific form of $A^{W}_{\text{closed}}(S,T)$. The relevant SVMPLs are always of the form (see \cite{Dixon:2012yy})
\begin{equation}
{\cal L}_W(z,\bar z)  = L_W(z) +L_{\tilde W}(\bar z) +   \text{product of lower weight}\,,
\end{equation}
where $\tilde W$ is $W$ with reversed letters. Then we can write
\begin{equation}
A^{W}_{\text{closed}}(S,T) = 2 i \sin (\pi S)  \sin (\pi T) \csc (\pi  (S+T)) I_2 \left( I_2^W+ I_2^{\tilde W} \right)+   \text{product of lower weights}
\end{equation}
In this form the symmetry $z \leftrightarrow \bar z$ is manifest and it is also clear that the holomorphic/anti-holomorphic factorisation of the type of 2d integrals relevant for the computation of the VS amplitude in AdS always involves combinations of the form $\left( I_2^W+ I_2^{\tilde W} \right)$. This does not appear to be an obvious symmetry of the Veneziano amplitude on $AdS$ found in this paper. 

\subsection{Explicit integrals}
Here we consider the explicit computation of $1d$ integrals, in terms of hypergeometric functions. We introduce the following notation
\begin{equation}
I[L_w(z)](S,T)= \int_{0}^1 dz (z)^{-S-1} (1-z)^{-T-1} L_w(z)\,,
\end{equation}
and consider insertions of growing complexity. When no insertions are present we have
\begin{equation}
I[1](S,T) = \frac{\Gamma (-S) \Gamma (-T)}{\Gamma (-S-T)}\,.
\end{equation}
Taking derivatives w.r.t.\ $S$ and $T$ we obtain the extra insertion of $\log z$ and $\log(1-z)$. Next we consider the insertion of 

\begin{equation}
 \text{Li}_n(z) =-L_{0^{n-1} 1}(z)\,.
\end{equation}
In this case we obtain
\begin{equation}
I[\text{Li}_n(z)](S,T) =\frac{\Gamma (-S+1) \Gamma (-T) }{\Gamma (-S-T+1)}\, _{n+2}F_{n+1}(1,\cdots,1,-S+1;2,\cdots,2,-S-T+1;1)\,.
\end{equation}
This in turns leads to 
\begin{equation}
I[\text{Li}_n(1-z)](S,T) =I[\text{Li}_n(z)](T,S)\,.
\end{equation}
Finally, as already mentioned
\begin{equation}
I[(\log z)^p (\log(1-z))^q \text{Li}_n(z)](S,T) =(-1)^{p+q} \partial^p_{S}\partial^q_{T}I[\text{Li}_n(z)](S,T) \,.
\end{equation}
This leads to an expression for any insertion that appears in our main result \eqref{G1_Li}.

\bibliographystyle{JHEP}
\bibliography{ads_veneziano}
\end{document}